\documentclass[%
 preprint,prx,
amsmath,amssymb,
superscriptaddress,
aps, 11pt]{revtex4-1}

\usepackage{amsmath}
\usepackage{amssymb}
\usepackage{dcolumn}
\usepackage{bm}
\usepackage{hyperref}
\usepackage{verbatim}
\usepackage{color}
\usepackage{xcolor}
\usepackage{mhchem}

\usepackage{verbatim}
\usepackage{graphics}
\usepackage{lipsum}% http://ctan.org/pkg/lipsum
\usepackage{graphicx}% http://ctan.org/pkg/graphicx

\usepackage{subfigure}
\usepackage{wrapfig}
\usepackage{epsfig}
\usepackage{float}
\usepackage{array}
\usepackage{psfrag}
\usepackage{color}
\usepackage{bbold}

\definecolor{azure}{rgb}{0.0, 0.5, 1.0}
\definecolor{asparagus}{rgb}{0.53, 0.66, 0.42}
\definecolor{ballblue}{rgb}{0.13, 0.67, 0.8}
\definecolor{sgreen}{rgb}{0.0, 0.8, 0.35}
\definecolor{darkgreen}{rgb}{0.0, 0.5, 0.0}
\definecolor{sred}{rgb}{0.9, 0.6, 0.4}
\usepackage{textcomp}

\usepackage{tikz}

\usepackage[]{color}

\begin{document}

%%%
\title{
%Theory of Active Emulsions in Membranes and 
%Wetting and Prewetting 
%Thermodynamics of Surface Phase Transitions with Membrane Binding
%
%Thermodynamics of Wetting and Prewetting with Membrane Binding
%
%How binding affects  bulk and surface condensation 
%
%Effects of surface binding on wetting and prewetting phase transitions
%
%submitted: Wetting and Prewetting Phase Transitions facilitated by Surface Binding
%Wetting, Prewetting and Surface Phase Transitions facilitated by Surface Binding
% 
 {Thermodynamics of Wetting, Prewetting and Surface Phase Transitions with
Surface Binding
}
%
%
%How binding affects wetting and prewetting phase transitions
%
%Wetting and prewetting phase transitions with surface binding
}
%%%

%%%%%%
\author{{Xueping Zhao}}
\affiliation{Max Planck Institute for the Physics of Complex Systems,
N\"{o}thnitzer Strasse~38, 01187 Dresden, Germany}
\affiliation{Center for Systems Biology Dresden,  Pfotenhauerstrasse~108, 01307 Dresden, Germany}
 
\author{Giacomo Bartolucci}
\affiliation{Max Planck Institute for the Physics of Complex Systems,
N\"{o}thnitzer Strasse~38, 01187 Dresden, Germany}
\affiliation{Center for Systems Biology Dresden,  Pfotenhauerstrasse~108, 01307 Dresden, Germany}
 
\author{Alf Honigmann}
\affiliation{Max Planck Institute of Molecular Cell Biology and Genetics, Pfotenhauerstrasse 108, 01307 Dresden, Germany}
\affiliation{Center for Systems Biology Dresden,  Pfotenhauerstrasse~108, 01307 Dresden, Germany}

%%%%%%
\author{Frank J\"ulicher$^a$}
\affiliation{Max Planck Institute for the Physics of Complex Systems,
N\"{o}thnitzer Strasse~38, 01187 Dresden, Germany}
\affiliation{Center for Systems Biology Dresden,  Pfotenhauerstrasse~108, 01307 Dresden, Germany}
\affiliation{Cluster of Excellence Physics of Life, TU Dresden, 01062 Dresden, Germany}

%%%%%%%%
\author{Christoph A.\ Weber
\footnote{\label{CA}Corresponding authors:  {julicher@pks.mpg.de} and 
 {christoph.weber@physik.uni-augsburg.de}}
}
\affiliation{Max Planck Institute for the Physics of Complex Systems,
N\"{o}thnitzer Strasse~38, 01187 Dresden,
Germany}
\affiliation{Center for Systems Biology Dresden,  Pfotenhauerstrasse~108, 01307 Dresden, Germany}

\date{\today}

%\tableofcontents

\begin{abstract}

In living cells, protein-rich condensates  can wet the cell membrane and surfaces of membrane-bound organelles.
Interestingly, many phase-separating proteins also bind to membranes leading to a molecular layer of bound molecules.
Here we investigate how binding to membranes affects    {wetting, prewetting and surface phase transitions. }
We derive a thermodynamic theory for a three-dimensional bulk in the presence of a two-dimensional, flat membrane.
 {At phase coexistence}, we find that membrane binding 
facilitates complete wetting and thus lowers the wetting angle.
Moreover, below the saturation concentration, binding facilitates the formation of a thick layer at the membrane and thereby shifts the prewetting phase transition far below the saturation concentration.
The distinction between bound and unbound molecules near the surface leads to a large variety of surface states  {and  complex surface phase diagrams with a rich topology of phase transitions.}
Our work suggests that surface phase transitions combined with molecular binding represent a versatile mechanism to control the formation of protein-rich domains at intra-cellular surfaces.
\end{abstract}

%\pacs{}% insert suggested ke/Users/admin/Matlab(Mac)R2010a segmentedywords - APS authors don't need to do this
%\keywords{}

\maketitle

%%%%%%%%%
%%%%%%%%%
%colors used for illustration:

% \tableofcontents

%===================================
%               Introduction
%===================================
\section{Introduction}

Surfaces introduce a new level of complexity, as Wolfgang Pauli alluded to in his famous quote:
``God made the bulk; surfaces were invented by the devil''~\cite{Pauli_quote_1999}.  {This complexity is ubiquitous in nature since surfaces are key determinants in many biological systems.}
A paradigm are living cells which are surrounded by a membrane and that contain intricate organelles enclosed by membranes.
While a major function of membranes is to compartmentalize biochemical reactions in cells, the interplay between membranes and the cellular bulk is crucial for many biological processes.
Examples are sensing and signalling, endocytosis
or asymmetric cell division.
%~\cite{Doherty&Harvey_Annu_rev_2009} %or asymmetric cell %division~\cite{knoblich2001asymmetric, Grill&Frank2019, Lars2019}.
%, to name just a few. % Not sure what type of references you want here, since the processes are so general. Maybe no references needed. This is textbook knowledge.
In addition to membranes, cells use protein-rich condensates to organize intra-cellular space. Such condensates  coexist with the cellular environment and share hallmark properties of physical droplets~\cite{Brangwynne2009, 
Hymann_Weber_Julicher_2014,
Banani&Hyman&Rosen&2017,Brangwynne_Hymann_PNAS_2011,Pappu_Brangwynne_Cell_2016}.
Interestingly, protein condensates can adhere to intracellular surfaces and membrane-bound organelles~\cite{Gall1999,
Brangwynne2009,
Knorr_wetting_nature_2021}, which resemble condensates wetted on surfaces. 
These observations indicate that phase-separating proteins in living cells not only phase-separate in the bulk but can also undergo  {phase transitions related to the membrane surface}. 

The theory of surface phase transitions was developed by Cahn~\cite{Cahn_wetting_1977}. In this seminal work, he discussed the graphical construction for wetting transitions but also 
showed the existence of prewetting transitions.
 {In the same year, Ebner and Saam reported wetting and prewetting transitions using density-functional theory~\cite{Ebner_Saam_PRL_1977}.}
The wetting transition separates the regime of complete wetting and partial wetting, where the interface of wetting droplets exhibits a contact angle with respect to the surface~\cite{Cahn_wetting_1977, Nakanishi_PRL_1982,deGennesreview1985, Diehl1985, PRB_Martin_1987, Diehl_1997, Daniel2009, Lipowsky_JPCB_2018, Dietrich_2020}.
While wetting solely occurs inside the binodal of the corresponding bulk phase diagram ( {coexistence} domain), prewetting phase transitions take place in the undersaturated regime where droplets shrink and disappear in the bulk.
Inside the prewetting regime of the phase diagram, a three-dimensional layer forms close to the surface
which we refer to as the ``thick layer''. 
The layer thickness can be quantified by the excess surface concentration which serves as an order parameter for the wetting and prewetting transitions.
Crossing the prewetting transition, e.g. by lowering the bulk concentration, the excess surface concentration (order parameter) decreases discontinuously to a lower value and molecules accumulate only very weakly at the surface, forming the  ``thin layer''. 
The close vicinity of the prewetting transition line and the binodal for many polymeric systems was an experimental challenge to distinguish bulk and surface transitions~\cite{Taborek1992, Chandavarkar1992, Bonn1992}.

Many of the phase-separating proteins have molecular domains by which they can bind to membranes~\cite{banjade2014phase, su2016phase, Oliver_Cell_2019}. Reversible membrane binding of proteins in the presence of chemical feedback is known to give rise to reaction-diffusion patterns on membranes~\cite{Loose_Schwille_Science_2008, Halatek_Cell_2012, Ramm_Schwille_2019}.
In contrast to wetted or prewetted states, these patterns are comprised of mono-layered, two-dimensional domains of specific protein composition leading to a spatially organized membrane.
Wetting or prewetting can however give rise to surface condensates that are three-dimensional. Such condensates can serve as hubs for down-stream assembly processes at membrane surfaces~\cite{Oliver_Cell_2019, Schwayer_Cell_2019, agudo2021wetting}.
Prewetting could also serve as a mechanism for the formation of proteins condensates on DNA strands~\cite{Quail2020.09.17.302299, Morin2020.09.24.311712} and thus play a role in chromatin organization. 
In summary, condensation at surfaces such as membranes or biofilaments appears as a key principle of spatial organization of biological surfaces~\cite{snead2019control}.
However,
a general thermodynamic theory that explores how  protein binding to  biological surfaces affects surface condensation is lacking.

In this work, we study  {the interplay between wetting, prewetting and surface phase transitions coupled via binding of molecules between bulk and membrane surface}.
We propose a general thermodynamic free energy that captures the molecular interactions in the three-dimensional bulk and a two-dimensional, flat surface.
Using this free energy, we determine the phase diagrams for wetting and prewetting at thermodynamic equilibrium.  By varying the parameters that describe surface binding of molecules, we find phase diagrams of variable complexity and topology which exhibit transitions between a rich set of thermodynamic surface states (Fig.~\ref{fig:phases_sketch}).
 {We find that binding 
can enlarge the regime of wetting and prewetting, and can move the prewetting transition to lower bulk concentrations. 
Another finding is that binding can give rise to prewetted states not only below the critical point but also in the absence of bulk phase separation.}

\begin{figure}
\centering
{\includegraphics[width=1\textwidth]{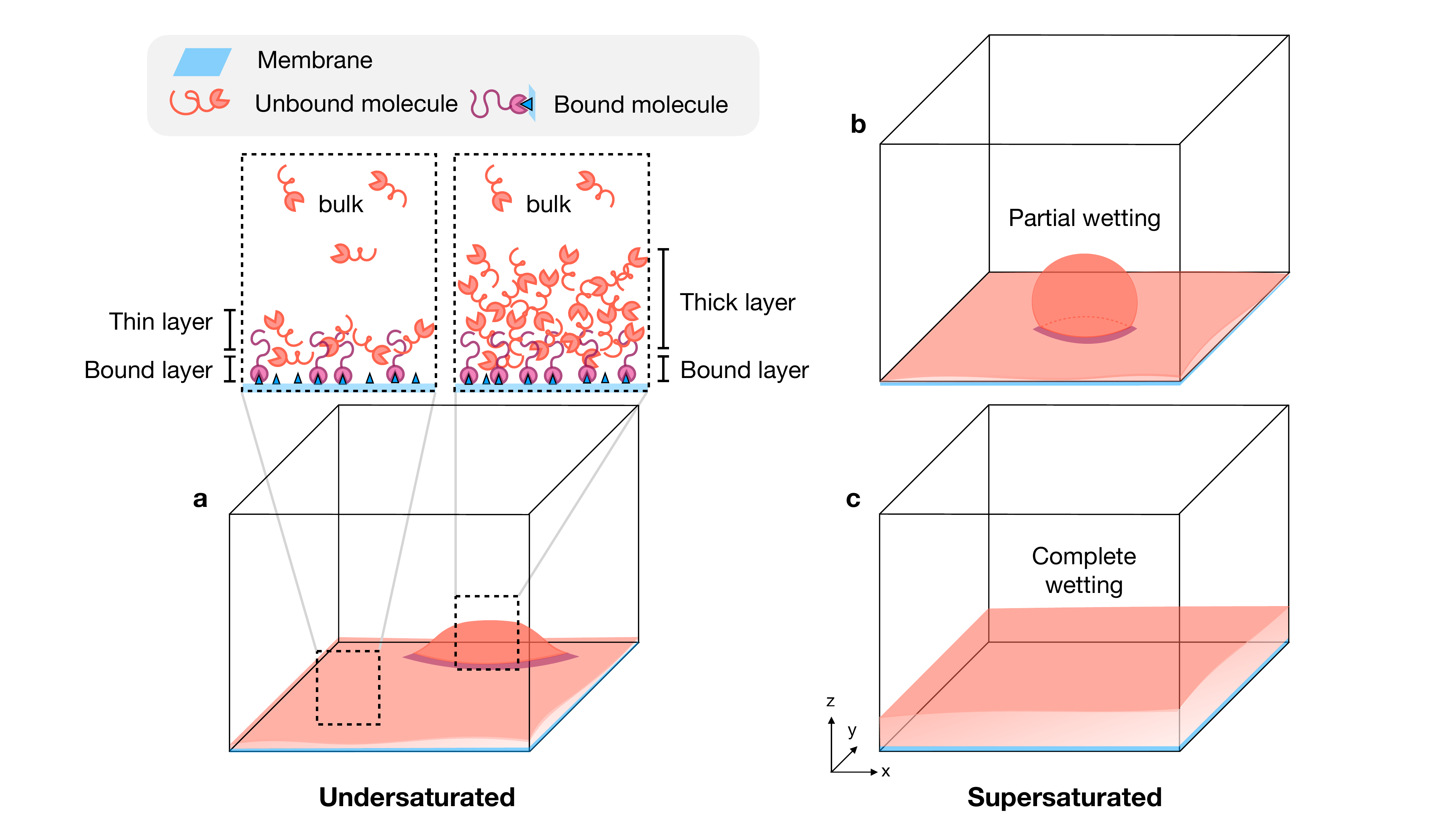}}
\caption{\textbf{Schematics of  phase transitions at surfaces in the presence of membrane binding}.
Molecules from the bulk can bind to specific sites (blue triangles) on the membrane (blue surface).
Unbound bulk molecules can also accumulate adjacent to the membrane surface, leading to the formation of three dimensional layers on the membrane surface (red surfaces). 
\textbf{(a)} If the system is undersaturated ($\phi_\infty<\phi_\text{out}^\text{eq}$), prewetted thin and thick layers can transiently form. At thermodynamic equilibrium, either thin or thick layers are stable except at the prewetting transition where both states coexist.  {At phase coexistence}, condensates either partially wet \textbf{(b)} or completely wet the membrane surface \textbf{(c)}, depending on the molecular interactions and the interactions with the surface.
 {In addition, bound molecules can phase-separate in the membrane, which is a surface phase transition.}
}
\label{fig:sketch}
\end{figure}

% \newpage

\section{Thermodynamics of phase separation with membrane binding}

We consider a liquid solution where solute molecules can bind to specific sites on a two-dimensional, flat membrane at $z=0$.
The volume fraction of molecules in the bulk is denoted by $\phi(x,y,z)$, the density of molecules bound to the membrane is described by the area fraction $\phi_m(x,y)$. During binding and unbinding events molecules transition between the solute state and the surface bound state according to
\begin{align}\label{eq:binding_reaction}
\phi \,  \rightleftharpoons  \phi_m \,  .
\end{align}
% DO NOT call it reaction scheme
 {Here we study how phase separation in bulk and surface affects  wetting and prewetting transitions by surface binding (Fig.~\ref{fig:sketch}a-c).}
To this end, we consider a bulk binary mixture 
of volume $V$ which is composed of solute molecules and solvent.
The free energy contains contributions from
the bulk $f_{b}(\phi)$, the membrane $f_m(\phi_m)$ and coupling free energy between them, $J(\phi|_{0}, \phi_m)$:
\begin{align}\label{eq:free_energy_F}
    F[\phi, \phi_m]  = \int_V d^3 {x} \Big [ f_{b}(\phi) + \frac{1}{2}\kappa |\nabla \phi|^2 \Big ]  +  \int_m d^2 {x} \Big [ f_m(\phi_m) + \frac{1}{2}\kappa_m |\nabla_{||} \phi_m|^2 + J(\phi|_{0}, \phi_m) \Big ] \, ,
\end{align}
where $\phi|_{0}$ is the bulk volume fraction at the membrane surface. Moreover, $\kappa$ and $\kappa_m$ characterize the corresponding free energy penalties for gradients in bulk and membrane, respectively, and  $\nabla_\parallel=(\partial_x,\partial_y)$ denotes the gradient vector in the membrane plane.
From Eq.~\eqref{eq:free_energy_F} we can define the local chemical potential $\mu=\nu_b \delta F/\delta \phi$ and $\mu_m= \nu_m \delta F/\delta \phi_m$, where 
 $\nu_b$ and $\nu_m$ denote the molecular volume and the molecular surface area of the molecules, respectively.

\subsection{Thermodynamics of a semi-infinite system}

We consider a semi-infinite, thermodynamic system with the membrane at $z=0$. Systems that are  homogeneous in the $x$-$y$ plane become effectively one-dimensional with a bulk volume fraction $\phi(z)$ changing along the $z$-direction with $z \in [0, \infty )$. 
The corresponding Helmholtz surface free energy functional reads
\begin{align}\label{eq:Helmholtz_energy}
    f_s[\phi,\phi_m] = \int_0^{\infty} dz \Big [ f_{b}(\phi) - f_{b}(\phi_{\infty})+ \frac{1}{2}\kappa |\partial_z \phi|^2 \Big ] +   f_m(\phi_m)  + J(\phi|_{0}, \phi_m) \, ,
\end{align}
where $\phi_\infty= \phi(z\to \infty)$ 
with the corresponding the external chemical potential
$\mu_\infty= df_b/d\phi|_{\phi=\phi_\infty}$.
We also define the excess surface concentration  
\begin{equation}
\label{eq:surface_excess_conc}
    c_s= %(\vect{x}_\parallel) =
    \int_0^{\infty}dz \left[ \frac{1}{\nu_b} \left(\phi(z) - \phi_\infty \right) \right] \, .
\end{equation} 
%at position $\vect{x}_\parallel=(x,y)$ on the surface.  
 {We obtain 
the surface free energy $f_s(c_s,\phi_m)$
when evaluating $f_s$ for the profile $\phi(z)$ that minimizes Eq.~\eqref{eq:Helmholtz_energy} for fixed $c_s$ and $\phi_m$, and with 
$\phi_\infty$ given far away from the membrane.}
The surface free energy $f_s(c_s,\phi_m)$
depends on the membrane area fraction $\phi_m$ and the
excess surface concentration $c_s$ and
has units of an energy per area. 
The chemical potentials in bulk and membrane can now be expressed as
\begin{subequations}
\begin{align}
\label{eq:mus}
    \mu&=\frac{\partial f_s}{\partial c_s}\, , 
    \\
    \quad \mu_m&=\nu_m \frac{\partial f_s}{\partial \phi_m} \, .
\end{align}
\end{subequations}
We can use a Legendre transformation to define the Gibbs surface free energy which is the surface thermodynamic potential in the ensemble where the chemical potentials are fixed:
\begin{align}\label{eq:Gibbs_energy}
    \gamma_s(\mu, \mu_m) = f_s(c_s, \phi_m) - \mu c_s - \mu_m \frac{\phi_m}{\nu_m} \, .
\end{align}
The conjugate variables to each of the chemical potentials $\mu$ and $\mu_m$ are the excess surface concentration $c_s$ and the area fraction $\phi_m$,
\begin{subequations}
\begin{align}
    c_s &= -\frac{\partial \gamma_s}{\partial \mu}\, ,\\
    \phi_m &= -\nu_m \frac{\partial \gamma_s}{\partial \mu_m} \, .
\end{align}
\end{subequations}
 {In the following, both variables serve as  order parameters for wetting, prewetting  and surface phase transitions. In particular,   $c_s$ characterizes the bulk layer adjacent to the surface while $\phi_m$ describes the state of the membrane.
}  

\subsection{Free energy minimization in a semi-infinite system}

We determine the equilibrium states via minimization of the Helmholtz surface free energy functional
$f_s[\phi,\phi_m]$, while keeping 
$c_s$ and $\phi_m$ fixed. 
This is achieved by minimizing the 
the Gibbs surface free energy functional
 $\gamma_s[\phi,\phi_m] = f_s[\phi,\phi_m] - \mu c_s - \mu_m {\phi_m}/{\nu_m}$,
 where $\mu$ and $\mu_m$ act as Lagrange multipliers to impose fixed $c_s$ and $\phi_m$, respectively.   
At this minimum, $\delta \gamma_s=0$, where
\begin{align}\label{eq:total_surface_energy}
\delta \gamma_s[\phi,\phi_m] &=  \int_0^{\infty} dz \left[ \left(\frac{\partial f_{b}}{ \partial \phi}  - \frac{1}{\nu_b}\mu  - \kappa \partial^2_z \phi \right) \delta \phi \right] + \kappa \frac{d \phi}{d z} \delta \phi\big|_0^{\infty}\\ \nonumber
& \quad +  \left[ \frac{\partial f_m}{\partial \phi_m}  + \frac{\partial J(\phi\vert_0, \phi_m)}{\partial \phi_m} - \frac{1}{
\nu_m} \mu_m \right ]\delta \phi_m + \frac{\partial J(\phi|_0, \phi_m)}{\partial \phi|_0} \delta \phi\big|_0 
\end{align}
is the functional variation of the Gibbs free energy functional $\gamma_s[\phi,\phi_m]$. 
This leads to the equilibrium conditions:
\begin{subequations}\label{eq:EQ_cond}
\begin{align}
   \frac{\partial f_{b}}{ \partial \phi}  - \frac{1}{\nu_b}\mu_{\infty}  - \kappa \partial^2_z \phi & = 0 \, ,
   \qquad z \in [0, \infty) \, , 
   \label{eq:EQ_conda}
   \\ 
 \phi(z)\big|_{z\to \infty} &= \phi_\infty \, ,
    %  \kappa \frac{d \phi}{d z}\Big|_{z=\infty} & = 0 \, ,
 \label{eq:EQ_condb}
   \\
   - \kappa \frac{d \phi}{d z}\Big|_{z=0} + \frac{\partial J(\phi, \phi_m)}{\partial \phi}\Big|_{z=0} &= 0 \, , 
   \label{eq:EQ_condc}\\
   \frac{\partial f_m}{\partial \phi_m}  + \frac{\partial J(\phi, \phi_m)}{\partial \phi_m} \Big|_{z=0}  - \frac{1}{\nu_m}\mu_{\infty} & = 0 \, , 
   \label{eq:EQ_condd}
\end{align}
\end{subequations}
where we used that binding between membrane and bulk (Eq.~\eqref{eq:binding_reaction}) is at thermodynamic  equilibrium:
$\mu_m = \mu_\infty$, where $\mu_\infty$ denotes the chemical potential of the reservoir at $z\to \infty$.
 {Note that at thermodynamic equilibrium, the gradient of the bulk profile $\phi(z)$ vanishes far away from the membrane,  ${d \phi}/{d z}|_{z \to \infty}  = 0$.}

%\mu &= \mu_\infty 
By integration over the bulk  (see Appendix~\ref{APP:graphical_construction})
and using Eq.~\eqref{eq:EQ_condb}, we can rewrite Eqs.~\eqref{eq:EQ_cond}:
\begin{subequations}\label{eq:finale_EQ_cond}
\begin{align}
\label{eq:finale_EQ_conda}
  \partial_z \phi  \pm \sqrt{\frac{2}{\kappa} W(\phi)} &=0, \qquad z \in [0, \infty) \, ,\\
 % \label{eq:finale_EQ_condb}
%     \kappa \frac{d \phi}{d z}\Big|_{\infty} & = 0\, , \\
     \label{eq:finale_EQ_condc}
    \sqrt{2 \kappa W(\phi|_0) }\pm \frac{\partial J(\phi, \phi_m)}{\partial \phi}\Big|_0 &= 0 \, , \\
    \label{eq:finale_EQ_condd}
  \frac{\partial f_m}{\partial \phi_m}  + \frac{\partial J(\phi, \phi_m)}{\partial \phi_m} \Big|_{z=0}  - \frac{1}{\nu_m}\mu_{\infty} & = 0 \, , 
\end{align}
where $W(\phi) \, V $ is the free energy needed to create a uniform fluid volume $V$ of composition $\phi$ from the reservoir of composition $\phi_{\infty}$~\cite{Cahn_wetting_1977},
% {NAME}
\begin{equation}\label{eq:def_W}
W(\phi) =  f_{b}\left(\phi\right) - \mu_\infty \frac{1}{\nu_b}\phi + \Pi_{\infty}   \, , 
\end{equation}
\end{subequations}
and $\Pi_{\infty}= -f_{b}(\phi_{\infty}) + \mu_\infty \phi_{\infty}/{\nu_b}$ is the osmotic pressure of the particle bath at $z\to \infty$.
Note that $\pm$ indicates that, in general, we need to solve Eq.~\eqref{eq:finale_EQ_conda} and Eq.~\eqref{eq:finale_EQ_condc} with both signs to obtain all solutions of Eq.~\eqref{eq:EQ_cond} (for a detailed discussion see Appendix~\ref{APP:graphical_construction}  {and Fig.~\ref{fig:graphic_method_both_signs}(a)}).

We use Eqs.~\eqref{eq:finale_EQ_condc} and \eqref{eq:finale_EQ_condd}
to obtain  the membrane area fraction $\phi_m$ and bulk volume fraction at the surface $\phi|_0$.
The spatial bulk profile $\phi(z)$ then follows from Eq.~\eqref{eq:finale_EQ_conda}. 
Using this profile we can compute the excess surface concentration $c_s$ (Eq.~\eqref{eq:surface_excess_conc}).
The thermodynamic control parameter is the chemical potential of the reservoir, $\mu_\infty$.
We also use the volume fraction of the reservoir $\phi_\infty$ as control parameter, which is equivalent to the average volume fraction $\bar \phi$ in the thermodynamic limit. 
We calculate the surface phase diagrams as a function of $\bar \phi$ and parameters characterising the interactions among the molecules (see Sect.~\eqref{sect:results}).  

\subsection{Surface phase diagrams obtained via graphical construction}

\begin{figure}
\centering
{\includegraphics[width=1.0\textwidth]{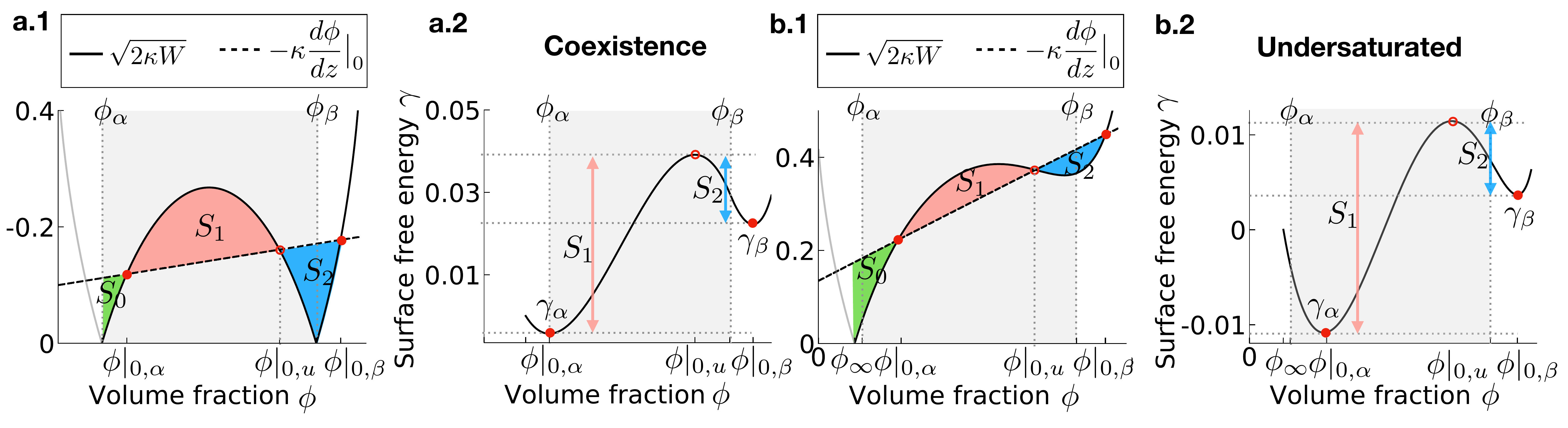}}
\caption{\textbf{Graphical construction for wetting and prewetting with membrane binding.} %\textbf{(a.1, b.1)}.
To obtain the equilibrium state at which the surface free energy $\gamma_s$ is minimal, we perform a graphical construction by comparing  $\sqrt{2\kappa W}$  (solid line) and $-{\partial \hat{J}(\phi, \phi_m)}/{\partial \phi} = - \kappa d\phi/dz\vert_0$ (dashed line) for wetting \textbf{(a.1)} and prewetting \textbf{(b.1)}, respectively.
The intersections of both functions can give two unstable solutions (open red circles) and two locally stable solutions (red dots). The locally stable solution that has the lower Gibbs surface free energy $\gamma_s$  corresponds to the thermodynamic equilibrium state; see \textbf{(a.2)} for wetting and \textbf{(b.2)} for the case of prewetting. 
%The shaded areas represent the energy gain ($S_1$, red) and loss ($S_2$, blue) when changing between different states with surface volume fraction $\phi\vert_{0,i}$, where $i = \alpha, \beta, u$.
%, i.e. $\gamma_{\alpha}$ + $S_1$ - $S_2$ =$\gamma_\beta$.
At a surface phase transition, the Gibbs surface free energies $\gamma_{\alpha} = \gamma(\phi|_{0, \alpha}, \phi_\infty)$ and 
$\gamma_{\beta} = \gamma_s(\phi|_{0, \beta},\phi_\infty)$
are equal (Eq.~\eqref{eq:transition_cond}), which amounts to $S_1=S_2$ (Eq.~\eqref{eq:gammas_S1S2}).
The gray shaded area represents the coexistence regime where the bulk can phase separation into a dilute phase ($\phi_{\alpha}$) and a dense phase ($\phi_{\beta}$).
Here, we only illustrate the branch $\sqrt{2\kappa\,W}$  {and depict the non-physical branch for $\phi<\phi\vert_{\alpha}$ in light grey}; see Fig.~\ref{fig:graphic_method_both_signs}(a) for the graphic construction with both branches $\pm \sqrt{2\kappa W}$. 
}
\label{fig:graphic_method}
\end{figure}

The transition lines separating different thermodynamic states at the surface can also be obtained via a graphical construction.
Using Eqs.~\eqref{eq:finale_EQ_conda}, we can rewrite the Gibbs surface free energy functional (Eq.~\eqref{eq:Gibbs_energy} with $f_s$ given by Eq.~\eqref{eq:Helmholtz_energy}) leading to 
\begin{align}\label{eq:Gibbs_energy_EQ}
   \gamma\left(\phi, \phi_\infty \right) 
   & = \int_{\phi_{\infty}}^{\phi} d\phi^\prime \Big[\pm \sqrt{2 \kappa W(\phi^\prime) } +
   \frac{\partial \hat{J}}{\partial \phi^\prime} 
   \Big] + \hat{J}\left(\phi_{\infty}, \phi_m \right)  \, ,
\end{align}
% \label{eq:gamma_s_Def}
%  xp{Do we need to change all the $\phi_{\infty}$ to $\bar{\phi}$ in the following? }\\
where $\hat{J}(\phi, \phi_m) =   J(\phi, \phi_m) + f_m(\phi_m) - {\nu_m^{-1}}\mu_\infty {\phi_m} $, with ${\partial \hat{J}}/{\partial \phi_m} = 0 $ (see Eq.~\eqref{eq:finale_EQ_condd}), and $\phi\vert_0$ and $\phi_m$ are determined by Eq.~\eqref{eq:finale_EQ_condc}.
%Moreover, $\hat{J}\left(\phi_{\infty}, \phi_m \right)$ is a constant with respect to all the equilibrium solutions. 
The integral term in Eq.~\eqref{eq:Gibbs_energy_EQ}  corresponds to
the area between $\pm \sqrt{2 \kappa W(\phi) }$ and $-{\partial \hat{J}}/{\partial \phi}$ and can be illustrated graphically (see colored areas in Fig.~\ref{fig:graphic_method}).

Now we can define the Gibbs surface potential as $\gamma_s=\gamma\left(\phi|_0, \phi_\infty \right)$.
Local extrema of the Gibbs surface potential correspond to the intersection points between $\pm \sqrt{2 \kappa W(\phi) }$ and $-{\partial \hat{J}}/{\partial \phi}$. There can be two 
local minima, 
$\gamma_\alpha=\gamma_s\left(\phi|_{0,\alpha}, \phi_\infty \right)$ and $\gamma_\beta=\gamma_s\left(\phi|_{0,\beta}, \phi_\infty \right)$, and the intermediate
local maximum is denoted by $\gamma_u=\gamma_s\left(\phi|_{0,u}, \phi_\infty \right)$.
The differences between these extremal values of the Gibbs surface potential can be expressed as the areas between $\pm \sqrt{2 \kappa W(\phi) }$ and $-{\partial \hat{J}}/{\partial \phi}$. Specifically, 
$\gamma_\alpha=-S_0 + \hat{J}\left(\phi_{\infty}, \phi_m \right) $, $\gamma_u=-S_0+S_1 + \hat{J}\left(\phi_{\infty}, \phi_m \right)$ and $\gamma_\beta=-S_0+S_1-S_2 + \hat{J}\left(\phi_{\infty}, \phi_m \right)$ (see colored domains in Fig.~\ref{fig:graphic_method}(a.1,b.1)). Thus, we find that the surface free energies at the minima are related by
\begin{align}
\label{eq:gammas_S1S2}
    \gamma_\alpha = \gamma_\beta + S_1 - S_2 \, .
\end{align}

At the prewetting and wetting transition lines, the Gibbs surface free energies of both states $\alpha$ and $\beta$ are equal:
\begin{align}\label{eq:transition_cond}
   \gamma_\alpha = \gamma_\beta \, ,
\end{align}
which implies, using Eq.~\eqref{eq:gammas_S1S2}, that $S_1 = S_2$ at the transition line.
This defines the graphical construction and determines the value of the control parameter, e.g.\ $\bar{\phi}$, at which the transition occurs. 
 {The minimized surface free energies exhibit a kink at both the wetting and prewetting transition (characterized by $c_s$) and the surface phase transition (characterized by $\phi_m$).
However, due to the coupling between bulk and membrane, both  order parameters, area fraction $\phi_m$ and the excess surface concentration $c_s$, in general change discontinuously at each of the transitions.}

% An illustration of the graphical construction for the case of wetting and prewetting is depicted in Fig.~\ref{fig:graphic_method}(a,b).

 {If the average volume fraction of the system $\bar{\phi}$ is within the domain of phase coexistence, i.e., $\phi_\alpha <\bar{\phi} <\phi_\beta$ for a certain range of interaction parameters, the homogeneous mixture is unstable and phase-separates into a dilute and a dense phase, with respective equilibrium values
$\phi_\alpha$ and $\phi_\beta$.}
Based on the definition of the Gibbs surface free energy density $\gamma_s=\gamma\left(\phi|_0, \phi_\infty \right)$ (Eq.~\eqref{eq:Gibbs_energy_EQ}),  {we identify the surface tensions between the membrane and the dilute phase $\gamma_{s,\alpha}$, 
between the membrane and the dense phase $\gamma_{s,\beta}$, and between the dilute and dense phase $\gamma_{\alpha, \beta}$, as follows:}
\begin{subequations}
\begin{align}
    \gamma_{s,\alpha} &=   \gamma(\phi\vert_{0, \alpha}, \phi_\alpha) 
  \, ,\\
     \gamma_{s,\beta} &=   \gamma(\phi\vert_{0, \beta}, \phi_\beta)  \, , \\
      \gamma_{\alpha, \beta} &= \int_{\phi_{\alpha}}^{\phi_{\beta}} d\phi \,  \sqrt{2 \kappa W } \, .
\end{align}
\end{subequations}
These equations satisfy the Young–Dupr\'e law, $ \gamma_{s, \alpha}= \gamma_{s, \beta}+\gamma_{\alpha, \beta} \, \cos(\theta)$, which defines the contact angle. The wetting transition is characterized by equal Gibbs surface free energy of the partially wetted state $\gamma_\alpha = \gamma_{s,\alpha}$, and the completely wetted state $\gamma_\beta = \gamma_{s,\beta}+ \gamma_{\alpha, \beta}$ (Eq.~\eqref{eq:transition_cond}),
 corresponding to zero contact angle, $\theta=0$.

 {If the average volume fraction of the system $\bar{\phi}$ is outside the domain of phase coexistence,
e.g., $\bar{\phi} < \phi_\alpha$ or $\bar{\phi} > \phi_\beta$,  there can still be two surface states corresponding to two local minima of the Gibbs surface free energy as can seen from the graphical construction, Fig.~\ref{fig:graphic_method}(b.2).
When Eq.~\eqref{eq:transition_cond} is satisfied and these free energies are equal, a phase transition occurs.
Due to the coupling between bulk and membrane, the corresponding phase transition in general  shares the characteristics of a
prewetting transition
with a discontinuous behavior in the excess surface concentration $c_s$ and a surface phase transition
with discontinuous behavior of the membrane area fraction $\phi_m$.
}

\subsection{Bulk and membrane free energies}

To calculate the surface phase diagram using the graphical construction as well as the profile $\phi(z)$ at thermodynamic equilibrium, we consider the following free energy for the bulk ($b$) and the membrane ($m$):
\begin{subequations}
\begin{align}
    f_b\left(\phi\right) &= \frac{k_B T}{\nu} \Big[ \frac{1}{n_b} \phi\, \ln \phi  +  (1-\phi)\, \ln \left(1-\phi\right) + \chi_b \phi\, (1-\phi) %+ \omega \phi 
    \Big]\, ,\\
       f_m(\phi_m) &= \frac{k_B T}{\tilde \nu} \Big[ \frac{1}{n_m} \phi_m\, \ln \phi_m + (1-\phi_m)\, \ln(1-\phi_m) + \chi_m \phi_m \,(1-\phi_m) + \omega_m \phi_m \Big]\, ,
\end{align}\label{eq:free_energy}
where $\nu$ and $\tilde \nu$ are the solvent molecular volume and solvent molecular surface area, respectively. 
Bulk molecules have a molecular volume of $\nu_b=\nu \, n_b$, while bound molecules occupy an area of $\nu_m=\tilde \nu \, n_m$. Here, $n_b$ and $n_m$ are the fractions of molecular volumes or surface areas of the molecules compared to the solvent, respectively.
 {Molecular interactions among molecules in the bulk and membrane are described by the interaction parameters $\chi_b$ and $\chi_m$, respectively.} Both parameters are in general different since molecular interactions can change upon binding. The internal free energy difference between membrane molecules and bulk molecules is denoted by $\omega_m$.

The coupling free energy between bulk and membrane reads: 
\begin{align}\label{eq:J_definition}
    J(\phi, \phi_m) = \frac{k_B T}{\tilde \nu} \Big[
    \omega_b\, \phi + {\chi}_{bb}\, \phi^2 + \chi_{bm} \, \phi \, \phi_m \Big] \, ,
\end{align}
\end{subequations}
where $\omega_b$ is the internal free energy of a bulk molecule at the surface. 
When the surface is attractive for molecules in the bulk, $\omega_b<0$.
 {The interaction parameter $\chi_{bb}$ accounts for enhanced interactions by accumulating  bulk molecules at the surfaces.
Note that in previous work,  wetting, prewetting and surface phase transitions were studied using the surface interaction parameters $\omega_b$ and $\chi_{bb}$~\cite{Cahn_wetting_1977, Nakanishi_PRL_1982}.}
 {In our model, we additionally account for the coupling between bulk and membrane surface via the parameter 
$\chi_{bm}$.
This parameter characterizes the interactions between membrane-bound molecules and bulk molecules at the surface. 
The term $\chi_{bm} \phi_m$ can also been considered as a further contribution  to the internal free energy $\omega_b$
due to molecules that are bound to the surface. (see Appendix \ref{APP:comparision_Nakanishi_Fisher} for details)
%For cases where the membrane bound fraction follows the bulk volume fraction ajacent to the surface, e.g. $\phi_m \propto \phi$, the coupling terms leads to a modified surface enhancement parameter $\chi_{bb}$ bridging our model to previous results~\cite{Cahn_wetting_1977, Nakanishi_PRL_1982}. 
}

% (see Appendix \ref{APP:comparision_Nakanishi_Fisher} for details)

%%%%%%%%%%%%%%%
 {\section{Effects of membrane binding on wetting, prewetting and surface phase transitions}
}\label{sect:results}

\subsection{Wetting and prewetting without phase separation in the membrane}\label{sect:results}

\begin{figure}[h!]
\centering
{\includegraphics[width=0.95\textwidth]{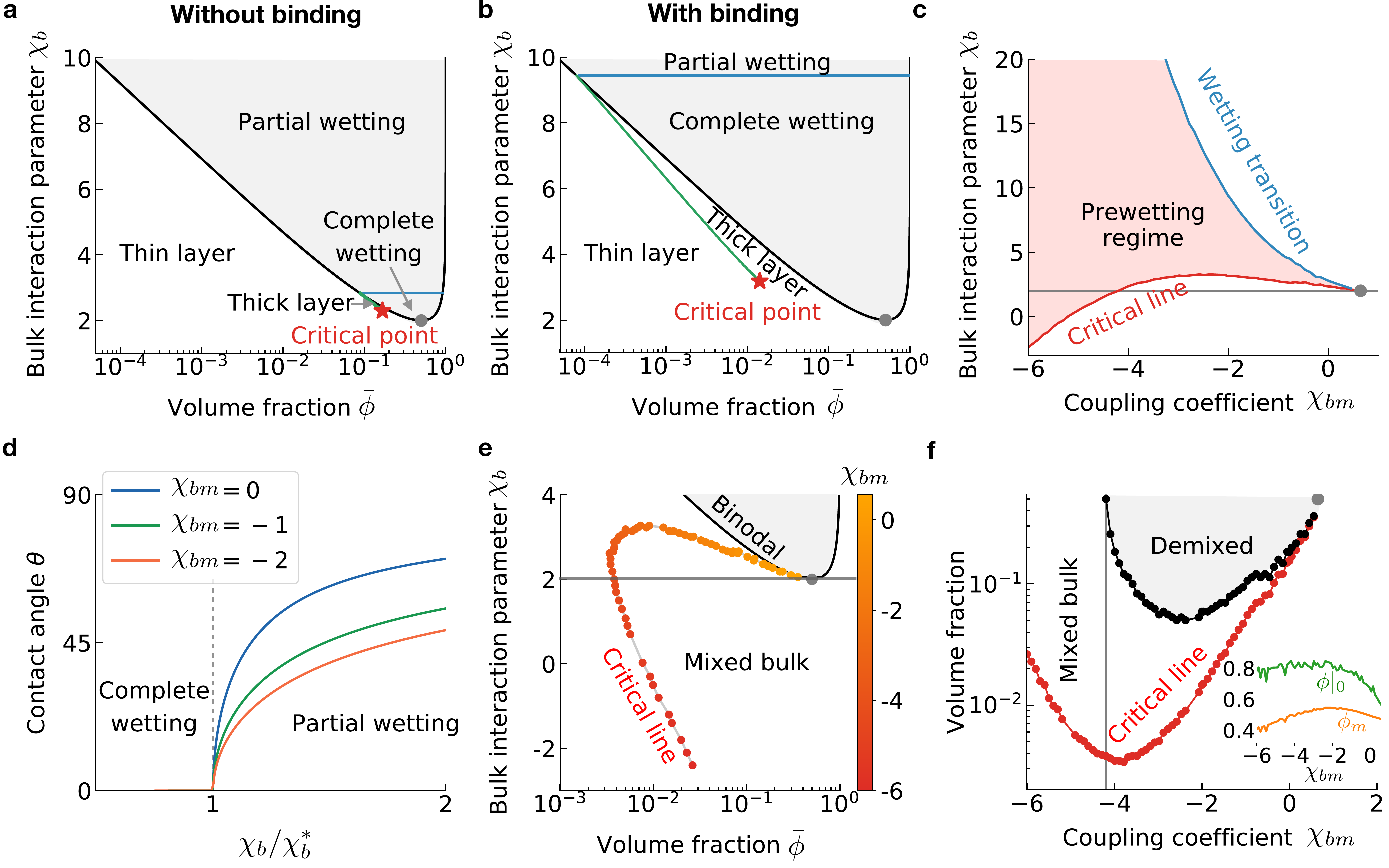}}
% Figure2.pdf
\caption{\textbf{Membrane binding  affects the wetting and prewetting.} \textbf{(a)} Surface phase diagram without binding to membrane. Note that ${\chi}_{bm} = 0$ leads to the same transition lines while heterogeneous states are different.
\textbf{(b)} Surface phase diagram with binding and attractive coupling between membrane and bulk (${\chi}_{bm} = -2$) illustrating that binding can shift the wetting transition line $\chi_b^*$ (blue) upwards and the prewetting
transition line (green) far away from the binodal (black). The red star is the prewetting critical point and the gray dot the bulk critical point.
\textbf{(c)} The prewetting regime (light red), enclosed by the wetting transition (blue) and the critical prewetting line (red),
widens as the coupling ${\chi}_{bm}$ gets more attractive.
\textbf{(d)} Contact angle $\theta$ of partially wetted condensates  {at phase coexistence} as a function of the rescaled bulk interaction parameter $\chi_b/\chi_b^*$ for
three values of the bulk-membrane coupling,
$\chi_{bm} = 0, -1, -2$.  This indicates that $\theta$ does not only change with $\chi_{bm}$ due to the shift of the wetting line. 
\textbf{(e, f)}  The prewetting critical line (shades of red) is  shifted to strongly undersaturated regimes for more attractive bulk-membrane coupling $\chi_{bm}$. There is a minimal critical volume fraction since
the critical $\chi_b$ decreases
for more attractive bulk-membrane coupling  $\chi_{bm}$.
Thus, 
molecules favor mixing with the bulk which is evident 
by a decrease of molecules adjacent and bound to the surface (see inset in (f)). Black dots in (f) represent the volume fractions of the binodal in (e) corresponding to the same $\chi_b$ values as the critical prewetting point. }
\label{fig:coupling_shift}
\end{figure}

Here, we first study how binding affects wetting and prewetting for cases where  {the surface free energy $f_m$ of membrane  cannot give rise to coexisting phases in the membrane} ($\chi_m=-4$).
% {In other words, there is no surface phase transition.}
For simplicity, 
the membrane insertion energy $\omega_m$ is set to zero.
Moreover, we consider  {an internal free energy corresponding to attractive interactions between membrane and bulk, $\omega_b = -0.3$, and no enhanced interactions,  $\chi_{bb}=0$}.
The effects of binding is studied for varying the bulk interaction parameter $\chi_b$, the bulk volume fraction  $\phi_\infty$, and the coupling parameter between bulk and membrane, $\chi_{bm}$.
 {We compare the corresponding results to the same system in the absence of membrane binding, $\chi_{bm}=0$.} Please note that the transition line for our model without binding ($\phi_m=0$) is the same as in the case with binding and vanishing bulk-membrane coupling ($\chi_{bm}=0$).
The corresponding surface phase diagram is shown in Fig.~\ref{fig:coupling_shift}a. This diagram depicts the domains with partially and completely wetted states separated by the wetting transition (blue line). It also shows the prewetting line (green line) separating thin and thick layer states.
At both transitions, the excess surface concentration $c_s$ is discontinuous.
The red star represents the prewetting critical point where thin and thick layer state become indistinguishable. 

\subsubsection{Membrane binding favors complete wetting and reduces the contact angle}

Binding to the membrane has significant effects on the wetting transition  (compare Fig.~\ref{fig:coupling_shift}a and~\ref{fig:coupling_shift}b).  
In particular, when increasing the attraction between the bulk and the membrane by reducing $\chi_{bm}$, complete wetting is favored.
This trend is evident in an upshift of the bulk interaction parameter at which the wetting transition occurs, $\chi_b^*$,  with decreasing coupling parameter $\chi_{bm}$ (blue line in Fig.~\ref{fig:coupling_shift}c).
Please note that at the wetting transition, there is a kink in the surface free energies  (Eq.~\eqref{eq:transition_cond}) and the excess surface concentration $c_s$ and membrane area fraction $\phi_m$ jumps.
In addition, for more attractive couplings between membrane and bulk, the wetting angle decreases (colored lines in Fig.~\ref{fig:coupling_shift}d). 
This behavior is not only due to the shift of the wetting transition $\chi_b^*(\chi_{bm})$.
Since the coupling is a second order term that describes the interactions between membrane and bulk, the trend of decreasing wetting angle with more attractive $\chi_{bm}$ persists even after recaling $\chi_b$ by the wetting transition $\chi_{bm}^*$.

\begin{figure}
\centering
{\includegraphics[width=1\textwidth]{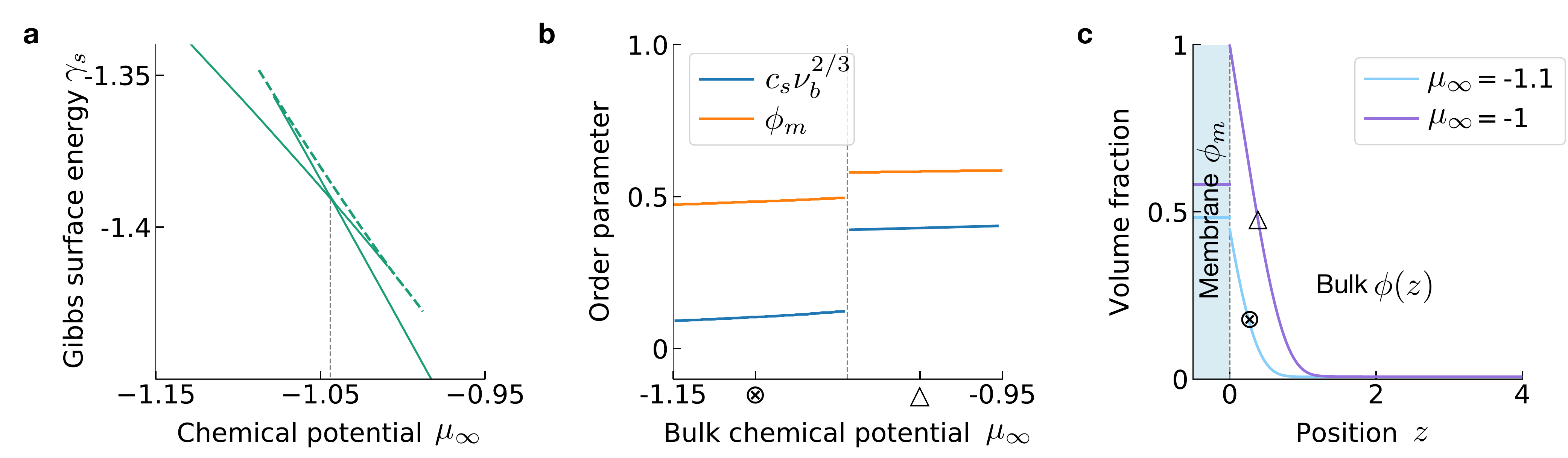}}
\caption{\textbf{Characteristics of prewetting phase transitions with binding and no phase separation in the membrane.} 
\textbf{(a)} The Gibbs surface energy $\gamma_s$ shows a kink at the prewetting phase transition ($\chi_b = 4$ in Fig.~\ref{fig:coupling_shift}b).  
\textbf{(b)} This kink in general implies a jump of both order parameters, the excess surface concentration $c_s$ and the  {area} fraction of  molecules bound at the membrane, $\phi_m$.
\textbf{(c)}  {At values of $\mu_\infty$ below and above the jump, the
membrane area fraction $\phi_m$ and
the bulk volume fraction at the membrane-bound layer $\phi|_0$ are significantly different and thereby the thickness of the bulk $\phi(z)$ changes.}
 }
\label{fig:coupling_shift2}
\end{figure}

\subsubsection{Prewetting transition shifts to lower concentrations}

Membrane binding not only affects the wetting transition line $\chi_b^*(\chi_{bm})$ but also changes the prewetting transition line (compare green line in Fig.~\ref{fig:coupling_shift}a with Fig.~\ref{fig:coupling_shift}b). 
These changes result from the fact that  at the prewetting line not only $c_s$ but also the area fraction of bound molecules $\phi_m$ is in general discontinuous (Fig.~\ref{fig:coupling_shift2}). 
In particular, since the prewetting line is linked to the wetting transition at the dilute branch of the binodal,
the upshift of the wetting line for more attractive coupling also moves the prewetting line to smaller volume fractions $\phi$. 
In addition, the critical point (red star) changes in a non-linear fashion
with the coupling strength (Fig.~\ref{fig:coupling_shift}c).
Both trends significantly widen the prewetting regime (light red area),
making the prewetting regime accessible for a broad range of interaction and coupling parameters.  
Interestingly,
when the attractive coupling parameters is varied within a physically meaningful range in the order of a few $k_BT$,
the volume fraction of the prewetting critical point can decrease by more than one order of magnitude  (Fig.~\ref{fig:coupling_shift}(e,f)). This implies that, due to binding, thick layers on the surface can already form via prewetting at bulk volume fraction far below the  {saturation concentration}  (compare red data points to binodal indicated by black line).
Surprisingly, there is a minimum of critical volume fraction of the prewetting transition (Fig.~\ref{fig:coupling_shift}(e,f)). This minimum arises because
the critical values of the bulk interaction parameter $\chi_b$ also decrease for strongly attractive  coupling strength $\chi_{bm}$.
A decreased bulk interaction parameter corresponds to interactions among bulk molecules becoming less attractive or even repulsive, in turn disfavoring the presence of bulk molecules adjacent to the membrane (inset, Fig.~\ref{fig:coupling_shift}f). This also decreases  the population of molecules bound to the surface $\phi_m$ (inset).
We conclude that the minimal value of the critical prewetting volume fraction arises from a competition of the energy gain of molecules being bound and adjacent to the surface (favored for attractive coupling strength $\chi_{bm}$) and the energy gain of molecules mixing with solvent in the bulk (favored by negative bulk interaction parameter $\chi_b$ corresponding to attractive solvent-bulk  interactions).

\subsubsection{Prewetting transition persists below the bulk critical point}

In the absence of membrane binding,  prewetting transitions can only occur for bulk interaction parameters above the bulk critical point where bulk phase separation is possible (gray dots are below the red star in Fig.~\ref{fig:coupling_shift}a). 
The attractive coupling between bulk and membrane enables situations where the prewetting critical point shifts to values of bulk interaction parameter below the bulk critical point, where the bulk cannot phase-separate for any volume fraction $\phi$ (see domains separated by gray line, entitled ``Mixed bulk" in  Fig.~\ref{fig:coupling_shift}(e,f)). In other words, binding mediates prewetting that is robust against concentration perturbations and that controls the formation of three-dimensional thick layers at surfaces, while phase separation in the bulk is suppressed.  
Similar findings were recently reported in experimental studies of functionalized surfaces and confirmed by corresponding Brownian dynamic simulations~\cite{Mognetti_2019}. 

\begin{figure}
\centering
{\includegraphics[width=1\textwidth]{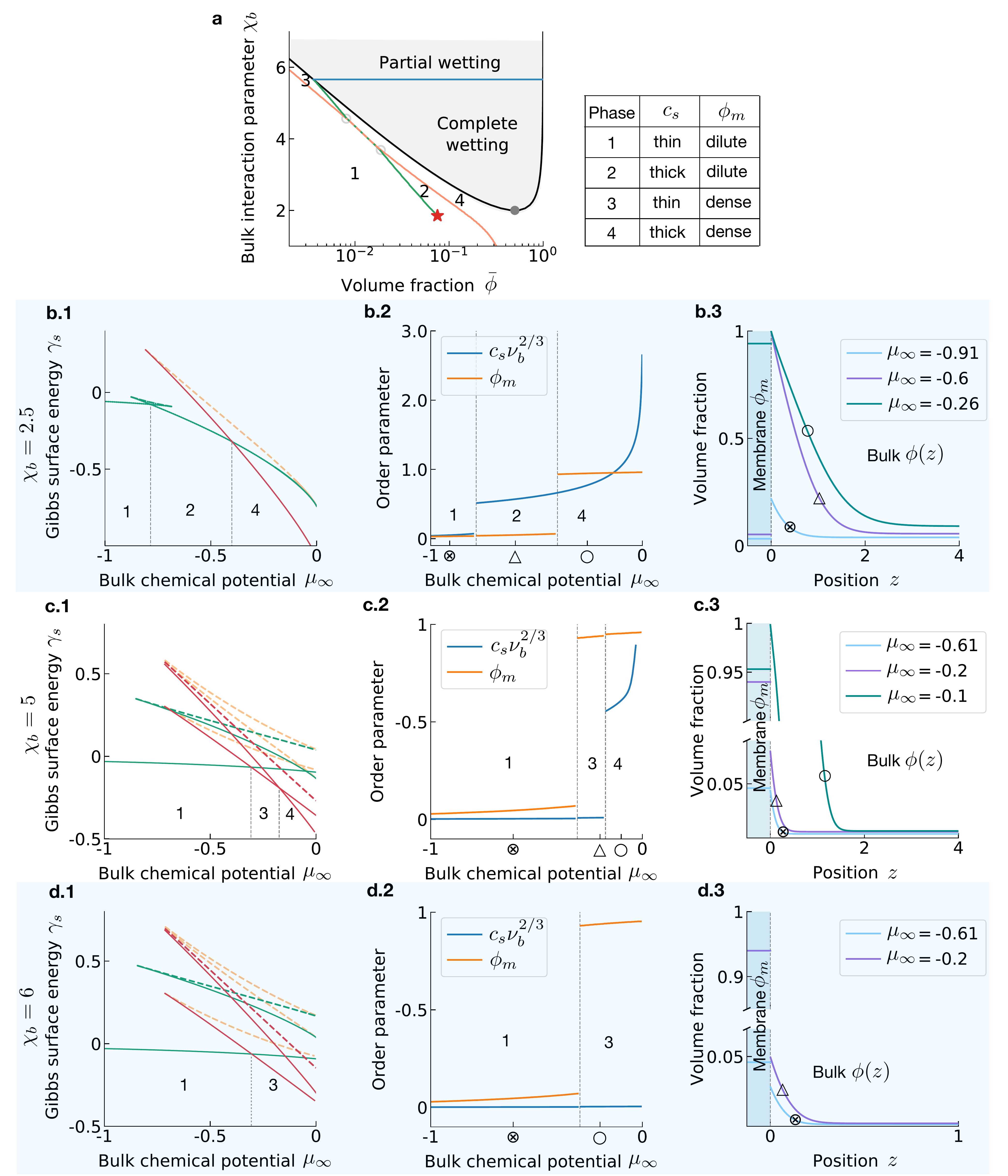}}
\caption{\textbf{Binding to phase-separated membranes 
with attractive bulk-membrane coupling.}
\textbf{(a)} 
Surface phase diagram showing multiple prewetting transition lines separating different surface states (see table). Parameter values are ${\chi}_{bm} = -0.1$ (attractive coupling), $\omega_m = -0.3$ (attractive membrane surface). We consider three bulk interaction values $\chi_b = 2.5, 5, 6$.
\textbf{(b.1, c.1, d.1)}
The Gibbs surface free energy $\gamma_s$ as a function of the chemical potential $\mu_\infty$.
%characterizing the amount of bulk molecules in the system.
%
\textbf{(b.2, c.2, d.2)} 
%The order parameters %characterizing the surface phase transition, i.e., the
Excess surface concentration $c_s$ (Eq.~\eqref{eq:surface_excess_conc}) and  membrane area fraction $\phi_m$.
\textbf{(b.3, c.3, d.3)}
Bulk volume fraction profiles $\phi(z)$  {and membrane area fraction} $\phi_m$.
}
 \label{fig:multiple_preweeting1}
 \end{figure}

\begin{figure}
\centering
{\includegraphics[width=1\textwidth]{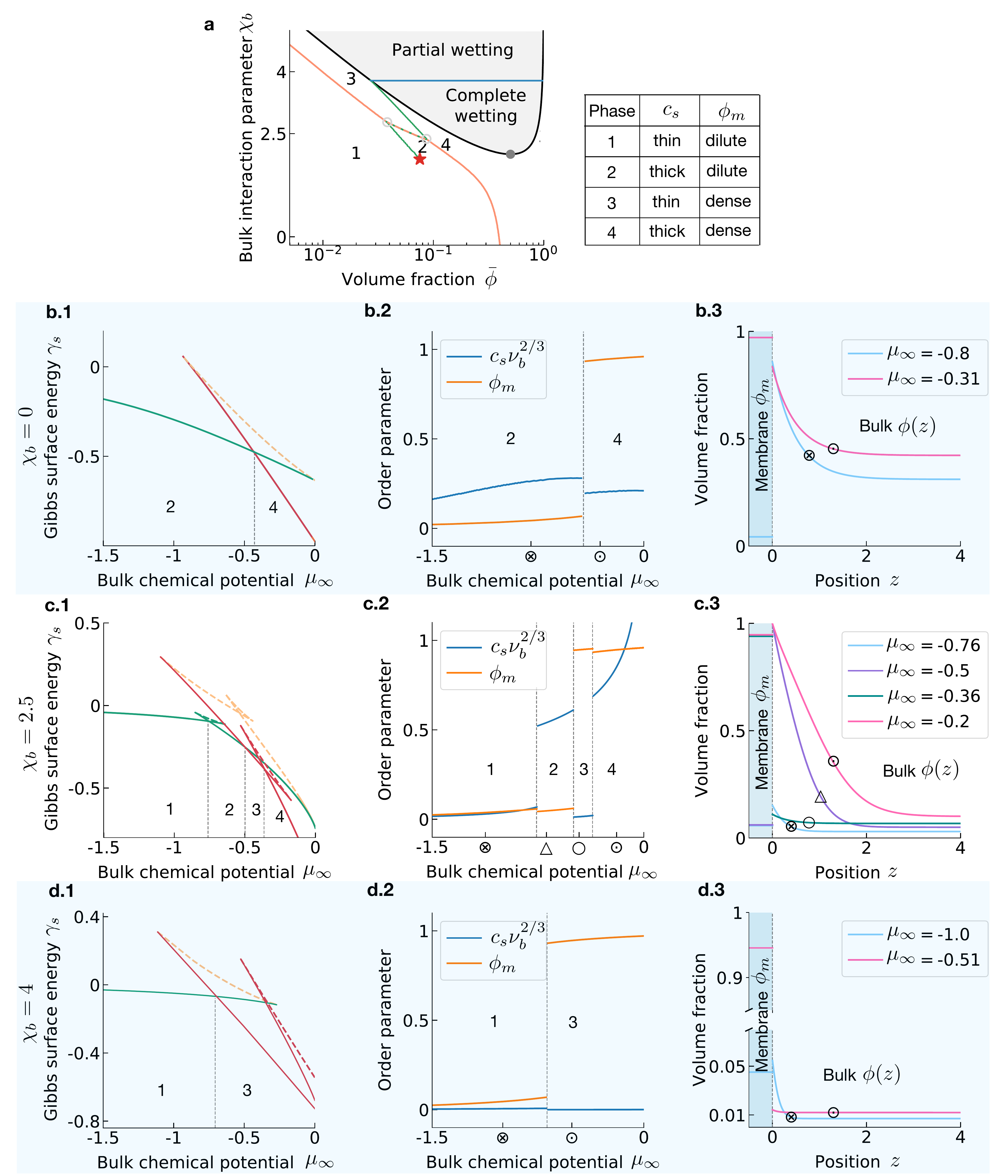}}
% Figure5.pdf
\caption{\textbf{Binding to phase-separated membranes 
with repulsive bulk-membrane coupling.} 
\textbf{(a)} Surface phase diagram showing multiple prewetting transition lines separating different surface states (see table);  ${\chi}_{bm} = 0.3$ and $\omega_m = -0.7$. 
We consider three bulk interaction values $\chi_b = 0, 2.5,$ and $4$.
\textbf{(b.1, c.1, d.1)}
The Gibbs surface free energy $\gamma_s$ as a function of the chemical potential $\mu_\infty$.
%characterizing the amount of bulk molecules in the system.
%
\textbf{(b.2, c.2, d.2)} 
%The order parameters %characterizing the surface phase transition, i.e., the
Excess surface concentration $c_s$ (Eq.~\eqref{eq:surface_excess_conc}) and  membrane area fraction $\phi_m$.
\textbf{(b.3, c.3, d.3)}
Bulk volume fraction profiles $\phi(z)$ and  {and membrane area fraction} $\phi_m$.
}
 \label{fig:multiple_preweeting2}
 \end{figure}

%%%%%%%%%%%%%%%
\subsection{Wetting and prewetting with phase separation in the membrane}\label{sect:results2}

%  {replace $\phi$ axes with $\phi_\infty$}\\

In this section, we will discuss how binding affects wetting and prewetting if molecules bound to the membrane can phase-separate.
 {In other words, the membrane can undergo a surface phase transition independent of the bulk,
which is realized by a positive interaction parameter $\chi_m>2$ that characterizes the interactions among membrane-bound molecules. In the following, we choose $\chi_m = 3$.}
In our studies, we fix the coupling coefficients characterizing the interactions of bulk molecules with the surface, $\omega_b = -0.3$ and $\chi_{bb} = -0.5$. 
We vary the average volume fractions $\bar{\phi}$ and the bulk interaction parameter $\chi_b$ to calculate the corresponding surface phase diagrams. We discuss two choices for the coupling strength between membrane and bulk: an attractive interaction parameter $\chi_{bm} = -0.1$ and  a repulsive one $\chi_{bm} = 0.3$, respectively. We find multiple striking effects on wetting and prewetting due to phase separation in membrane.

\subsubsection{ {Phase transitions} between four distinct surface states}

When bound molecules can phase-separate in the membrane, wetting and prewetting transitions are affected due to the mutual coupling between membrane and bulk.  
This coupling is characterized by the coupling coefficient $\chi_{bm}$, which links the behavior of the two respective order parameters $\phi_m$ and $c_s$.
In general, we find that there are four types of thermodynamic states   {outside the domain of phase coexistence}; see Fig.~\ref{fig:multiple_preweeting1}a and \ref{fig:multiple_preweeting2}a. 
These four surface states
are the combinatoric possibilities between a thick or thin surface layer (high or low $c_s \nu_b^{2/3}$), and a dense or dilute phase of membrane-bound molecules (high or low $\phi_m$).
Similar to the case without phase separation of bound molecules in the membrane (Fig.~\ref{fig:coupling_shift}c),
such new thermodynamic prewetted states can be found in a broad range of bulk volume fractions and bulk interaction parameters.

 {In the surface phase diagrams, the four surface states are separated by first order phase transition lines at which, in general, both order parameters $c_s$ and $\phi_m$ are discontinuous. The  corresponding phase transitions have mixed characteristics between a surface phase transition and a prewetting transition since bulk and membrane  are coupled.
However, we can test whether the transition lines exist for a decoupled bulk and membrane ($\chi_{bm}=0$), and thus label surface phase transition lines (orange) and prewetting lines (green), separately.
Due to the coupling, there can be phase transition lines that solely exist for a coupling between bulk and membrane  (orange-green-dashed). Transition lines that intersect at triple points (light grey circles) at which three different surface states coexist; see Fig.~\ref{fig:multiple_preweeting1}a and Fig.~\ref{fig:multiple_preweeting2}a.}

% discuss Gibbs loops
The existence of multiple thermodynamic surface states can lead to complex Gibbs loops around the transition between two thermodynamic states.
A classical Gibbs loop
consists of two locally stable branches, which are connected by a locally unstable branch.
To illustrate the Gibbs loop for prewetting transitions with membrane binding, we show the Gibbs surface energy $\gamma_s$ as a function of the chemical potential of the reservoir, $\mu_\infty$ (Fig.~\ref{fig:multiple_preweeting1}(b1,c1,d1) and~\ref{fig:multiple_preweeting2}(b1,c1,d1)).
At the phase transitions, the Gibbs surface energy of the thermodynamic state exhibits a kink, while the locally stable (solid lines) and unstable (dashed lines) branches form the Gibbs loops. 
We find that the complexity of such Gibbs loops is different for different values of the bulk interaction $\chi_b$. 
For example, we find classical Gibbs loops where two locally stable branches and one unstable branch exists in a certain chemical potential range of $\mu_\infty$  (Fig.~\ref{fig:multiple_preweeting2}(b1)).
We also find cases where up to four locally stable surface states exists 
(Fig.~\ref{fig:multiple_preweeting1}(b1,c1,d1) and~\ref{fig:multiple_preweeting2}(c1,d1)).
This structure of nested Gibbs loops suggest a complex dynamics towards equilibrium.

%%%%%%
\subsubsection{Correlated and anti-correlated jumps of surface order parameters}

%The coupling between membrane and bulk implies that in general both order parameters, the excess surface concentration $c_s$ and the bound area fraction in membrane $\phi_m$, jump at the transition lines. 
 {The sign of the discontinuity of both order parameters at the transition line} is determined by whether the coupling is attractive ($\chi_{bm}<0$) or repulsive ($\chi_{bm}>0$).
Specifically, for attractive couplings, both order parameters jump upwards as the chemical potential $\mu_\infty$ is increased, while for repulsive couplings, order parameters can show jumps in opposite directions; compare e.g. Fig.~\ref{fig:multiple_preweeting1}(c.2) and 
Fig.~\ref{fig:multiple_preweeting2}(c.2).

For attractive bulk-membrane couplings ($\chi_{bm}<0$), and increasing chemical potential $\mu_\infty$, either $c_s$ or $\phi_m$ shows a pronounced jump upwards.
This depends on the bulk interaction $\chi_b$ (compare Fig.~\ref{fig:multiple_preweeting1}(b.2) for $\chi_b=2.5$ with Fig.~\ref{fig:multiple_preweeting1}(b.3) for $\chi_b=5$). 
For more attractive bulk interactions ($\chi_b=5$ versus $\chi_b=2.5$), molecules prefer binding to accumulating at the membrane surface. Thus, the order parameters for membrane binding $\phi_m$ show the pronounced jump first as the chemical potential $\mu_\infty$ is increased (c.2), while for less attractive bulk interactions, the excess surface concentration $c_s$ makes the bigger jump first  (b.2). 
%To illustrate the different hierarchies of the order parameter jumps, we colorize the transition line according to a dominant $c_s$ jump (green line), a dominant $\phi_m$ jump (orange line) {need to be updated} in Fig.~\ref{fig:multiple_preweeting1}(a) and
%Fig.~\ref{fig:multiple_preweeting2}(a).
Thus, for the case of attractive bulk-membrane coupling, the layer width of the profile $\phi(z)$ as well as $c_s$ increases for increasing $\mu_\infty$ (Fig.~\ref{fig:multiple_preweeting1}(b.3,c.3, d.3)). In other words, the more molecules are in the bulk, the thicker the prewetted layer.

This behavior changes for the case of a repulsive bulk-membrane coupling ($\chi_{bm}>0$).
In this case, 
at each transition, both order parameters jump into opposite directions. 
Such opposite jumps can even cause  a decrease in layer thickness of the profile  $\phi(z)$ (Fig.~\ref{fig:multiple_preweeting2}(c.3)). 
Similar to the case of attractive couplings,
the jump heights and the number of jumps depend on the bulk interactions $\chi_b$ (Fig.~\ref{fig:multiple_preweeting2}(b.2, c.2, d.2)).

%%%%%%%%%%%%%%%
 {\subsection{Wetting, prewetting and surface phase transitions with related bulk and surface interactions}}\label{sect:results3}

In the last two sections we discussed  {wetting, prewetting and surface phase transitions} for bulk interaction parameters $\chi_b$ and coupling parameters between bulk and membrane $\chi_{bm}$ that were varied independently from each other. 
However, if molecules that are bound to the membrane are the same as the molecules in the bulk, both parameters describe similar physical interactions and are therefore related. 
This relationship is specific to the system of interest and can depend on
the membrane composition and the type of solvent and bulk molecule. 
To account for a relation between the interaction parameters $\chi_b$  and $\chi_{bm}$,
we consider for simplicity a linear relationship,
\begin{equation}\label{eq:chibm_chib}
    \chi_{bm} = -\alpha \, \chi_b \, ,
\end{equation}
where $\alpha$ describes the correlation between both interaction parameters. Interactions among bulk molecules and interaction of bulk with membrane-bound molecules
can have equal ($\alpha > 0$) or opposite ($\alpha < 0$) signs, which we refer to the correlated and anti-correlated case, respectively. 
A positive $\alpha$ applies if interactions in the bulk are similar to interactions between bulk and bound molecules. 
A negative $\alpha$ could for example correspond to a situation where the molecular domain mediating the interactions in the bulk $\chi_b$ is also involved in binding to the surface and thus not accessible for interactions of bound molecules with bulk molecules $\chi_{bm}$.
 {For simplicity, the interaction parameter describing interactions among membrane-bound molecules  $\chi_m$ is kept constant.}

\begin{figure}
\centering
{\includegraphics[width=1\textwidth]{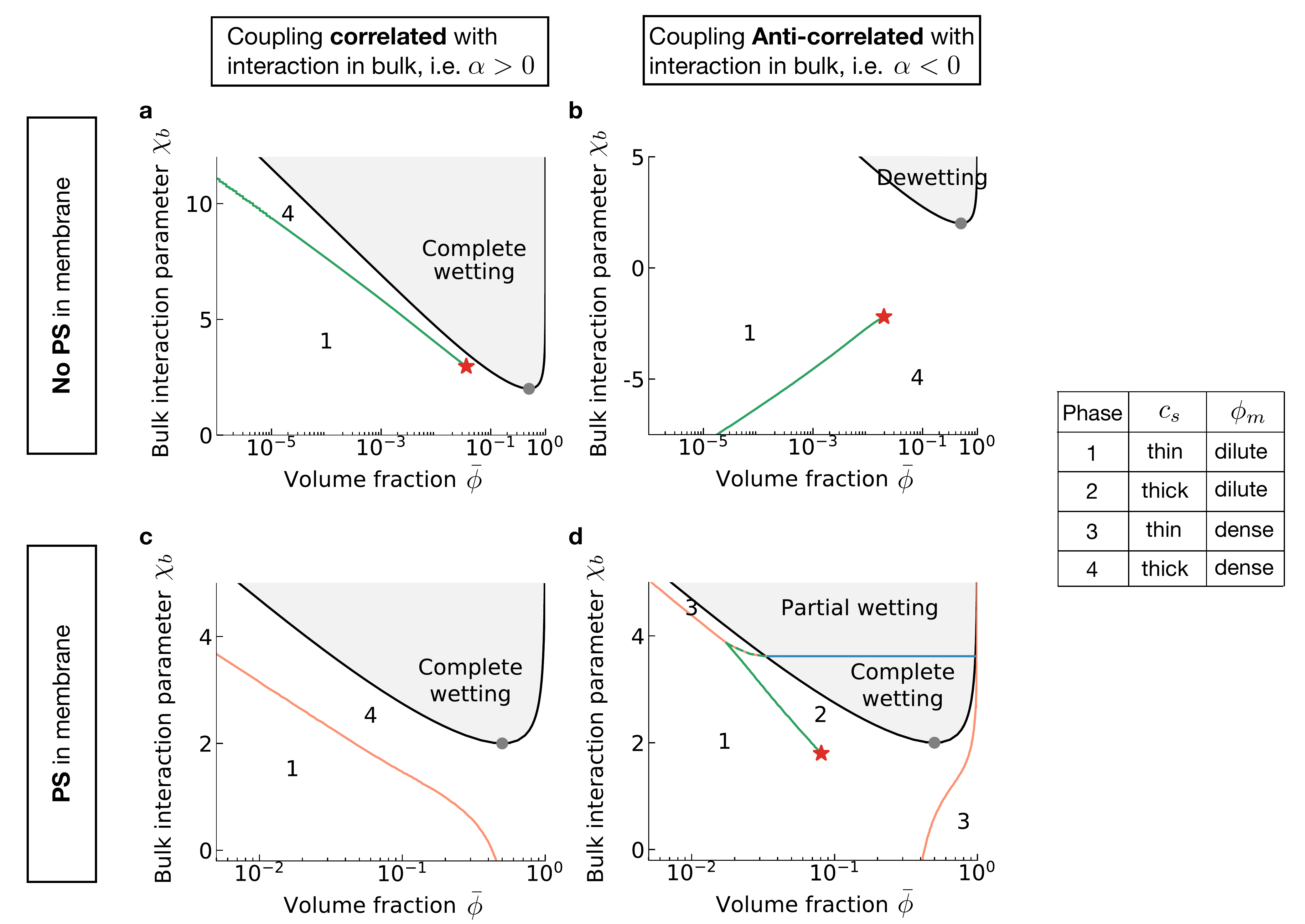}}
% slaved_phase_diagram.pdf
\caption{ {\textbf{Surface phase diagram with related interaction parameters $\chi_{bm}$ and $\chi_b$.} 
The membrane-bound molecules cannot phase-separate in \textbf{(a)} and \textbf{(b)} ($\chi_m = -4$), while they are able to phase-separate in \textbf{(c)} and \textbf{(d)} ($\chi_m = 3$). 
In addition, we distinguish relationships between bulk interactions $\chi_b$ and membrane-bulk interactions $\chi_{bm}$
of equal 
(correlated, $\alpha>0$) and opposite signs (anti-correlated, $\alpha<0$); see Eq.~\eqref{eq:chibm_chib}.
The basic parameter values are:
{(a)}  {$\omega_b = -0.3$, $\chi_m = -4$}, $\alpha = 0.4$.  {(b)}  {$\omega_b = -0.3$, $\chi_m = -4$},  $\alpha = -2.8$. {(c)}  {$\omega_b = -0.3$, $\chi_{bb} = -0.5$,  $\chi_m = 3$, $\omega_m = -0.3$}, $\alpha = 0.5$.  {(d)}  {$\omega_b = -0.3$, $\chi_{bb} = -0.5$,  $\chi_m = 3$, $\omega_m = -0.3$}, $\alpha = -0.5$. 
(\textbf{a},\textbf{b})
The prewetting transition line (green) is not connected  to the coexistence line, which is different from the case in models without surface binding~\cite{Cahn_wetting_1977, Nakanishi_PRL_1982}. 
{\textbf{(c)}} The surface phase transition (orange) can be either detached from the binodal,  or
{\textbf{(d)}} connected to  at the wetting transition.}
}
 \label{fig:slaved_phase_diagram}
 \end{figure}

In the following, we study the phase diagrams of wetting, prewetting and surface phase transitions for correlated and anti-correlated interaction parameters. In addition, we distinguish between the cases with ($\chi_m=3$) and without ($\chi_m=-4$) phase separation in the membrane. 
 {We find that the relation between interaction parameters $\chi_{bm}$ and $\chi_b$ (Eq.~\eqref{eq:chibm_chib}) can lead to a rich plethora of phase diagrams with complex topologies of phase transition lines that qualitatively differ to results obtained from models without surface binding~\cite{Cahn_wetting_1977,Nakanishi_PRL_1982}. For example, when molecules can bind to surfaces, prewetting transition lines can disconnect from the coexistence line, in particular from the wetting transition line (Fig.~\ref{fig:slaved_phase_diagram}(a,b)). Thus, 
prewetted states can occur for a broad range of bulk interaction parameters $\chi_b$.
Furthermore, the wetting transition is suppressed such that either complete wetting (Fig.~\ref{fig:slaved_phase_diagram}(a)) or dewetting  (Fig.~\ref{fig:slaved_phase_diagram}b) occurs inside the coexistence domain. 
Strikingly, the surface phase transition line is connected to the wetting
transition (Fig.~\ref{fig:slaved_phase_diagram}d).
This implies that 
phase separation in the membrane-bound layer can control the transition between partial and complete wetted states  (Fig.~\ref{fig:slaved_phase_diagram}d).}

\subsubsection{Suppression of  {wetting transition}}

We first discuss the case of  membrane-bound molecules that do not phase-separate in the membrane (e.g. $\chi_m = -4$).
For correlated $\chi_b$ and $\chi_{bm}$ ($\alpha>0$), partially wetted states are suppressed and complete wetting always occurs  {at coexistence regime} (Fig.~\ref{fig:slaved_phase_diagram}a). 
In this case, there is no wetting transition line and the prewetting line does not merge with the binodal.
Moreover, the prewetting region where thick layers form enlarges for increasing bulk interaction parameter $\chi_b$. 
The reason for the homogeneous states (completly wetted, thick layer) being favored on the membrane surface is that larger $\chi_b$ also implies more attractive bulk-membrane coupling ($\alpha>0$).
For anti-correlated $\chi_b$ and $\chi_{bm}$ ($\alpha<0$), wetting can be completely suppressed for attractive bulk interactions ($\chi_b>0$); see Fig.~\ref{fig:slaved_phase_diagram}b. In other words, within the  {coexistence} regime, there are no condensates at the membrane surface. However, prewetted states can form for repulsive interactions among bulk molecules ($\chi_b<0$).
This case represents an ideal scenario to either prevent bulk condensates from interacting with surfaces or to  
enable prewetted condensates at surface without the capability of the bulk to form condensates.

Now we discuss the case where  membrane-bound molecules can phase-separate in the membrane (e.g.  {$\chi_m = 3$}).
If $\chi_b$ and $\chi_{bm}$ are correlated ($\alpha>0$), complete wetting
is favored over 
partial wetting in the  {coexistence} regime (Fig.~\ref{fig:slaved_phase_diagram}c), similar to the case without phase separation in the membrane (Fig.~\ref{fig:slaved_phase_diagram}a).
Also, the domain of  {surface} states broadens as the bulk interaction parameters get more attractive. 
The only qualitative difference to the case without phase separation in the membrane is that there is no critical point of prewetting (Fig.~\ref{fig:slaved_phase_diagram}c). 
The reason is that though both, bulk interactions (negative $\chi_b$) and 
membrane-bulk interactions get more repulsive, the ability to phase-separate in the membrane can still enable thick layers at the membrane surface. 

\subsubsection{ {Phase separation in membrane facilitates  wetting}}

 {With independent interaction parameters $\chi_{b}$ and $\chi_{bm}$ (see Fig.~\ref{fig:multiple_preweeting1}a and Fig.~\ref{fig:multiple_preweeting2}a), the surface phase transition (orange lines) does not intersect with the binodal while the prewetting transition line (green lines) merges tangentially.
The latter is also the case in model without membrane binding~\cite{Cahn_wetting_1977, Nakanishi_PRL_1982}.
% Can we say where the tangential behavior comes from physically?
However, for anti-correlated $\chi_b$ and $\chi_{bm}$ ($\alpha<0$) and phase separation in the membrane (Fig.~\ref{fig:slaved_phase_diagram}d),
the surface phase transition intersects with the binodal line at the wetting transition. This finding indicates that  membrane-bound molecules coupled to the bulk can alter the interplay between wetting and prewetting  transition. 
%In other words, the wetting transition does not occur  due to the bulk internal free energy $\omega_b$ and surface enhancement parameter $\chi_{bb}$ as in the case of classical wetting/prewetting 
%[XXX Please check what happens to 7d for $\omega_b=\chi_{bb}=0$ ]. 
%Interestingly, it merges with binodal in a non-tangential way.
} 
Furthermore, 
for strongly attractive bulk interactions (large $\chi_b$) prewetting can even occur at conditions below the bulk critical point mediated by phase separation in the membrane.  
Finally, the membrane phase transition in the absence of bulk interactions (e.g. $\chi_b=0$) is now disconnected from the prewetting lines but converges to the binodal of the dense phase (Fig.~\ref{fig:slaved_phase_diagram}d).

% {@Xueping: I suppose that there can be a dewetting domain for larger $\chi_b$ ... let us discuss}

%When the membrane-bound molecules have a strong interaction e.g. $\chi_m = 3$, with strengthened bulk-membrane attractive interactions (i.e. $\alpha > 0$), freely diffusive molecules in the bulk is much easier to bind to membrane and thus promotes the phase separation (orange transition line) in the membrane (see Fig.~\ref{fig:slaved_phase_diagram}c). Moreover, order parameters $c_s$ and $\phi_m$ correlated with each other due to the attractive coupling $\chi_{bm} < 0$. Thus, thick layer forms after phase separation in the membrane (orange curve), and complete wetting condensates in the following bulk demixed region. Due to this attractive bulk-membrane interaction $\chi_{bm}$ would be even stronger at lager bulk-bulk interaction $-\chi_b$, wetting transition is suppressed in this case either. 

\subsubsection{Antagonism between membrane-bound and prewetted layers}

In classical wetting and prewtting  (Fig.~\ref{fig:coupling_shift}a),  the  profile of a thick layer curves upwards when approaching the surface due to attractive interactions between bulk and surface ($\omega_b<0$).
In the presence of membrane binding this leads to a high  {area} fraction $\phi_m$ in the membrane for correlated interaction parameters ($\alpha >0$).
This situation can fundamentally change when membrane binding and prewetting are antagonistic due to anti-correlated interaction parameters ($\alpha <0$).
As a result of this anti-correlation, we can distinguish between two types of antagonistic cases.
First, a thick prewetting layer near the surface
can induce detachment of molecules from the surface (purple lines in Fig.~\ref{fig:graphic_method_both_signs} {d}). 
Molecules cannot bind to the surface and thus the surface is effectively attractive. 
Second, 
molecules bind to the membrane surface which thereby becomes repulsive for bulk molecules.
No prewetting layer forms and the concentration profile  even curves downward when approaching the surface (see Fig.~\ref{fig:graphic_method_both_signs} {d}).

\section{Discussion}

\begin{figure}
\centering
{\includegraphics[width=1\textwidth]{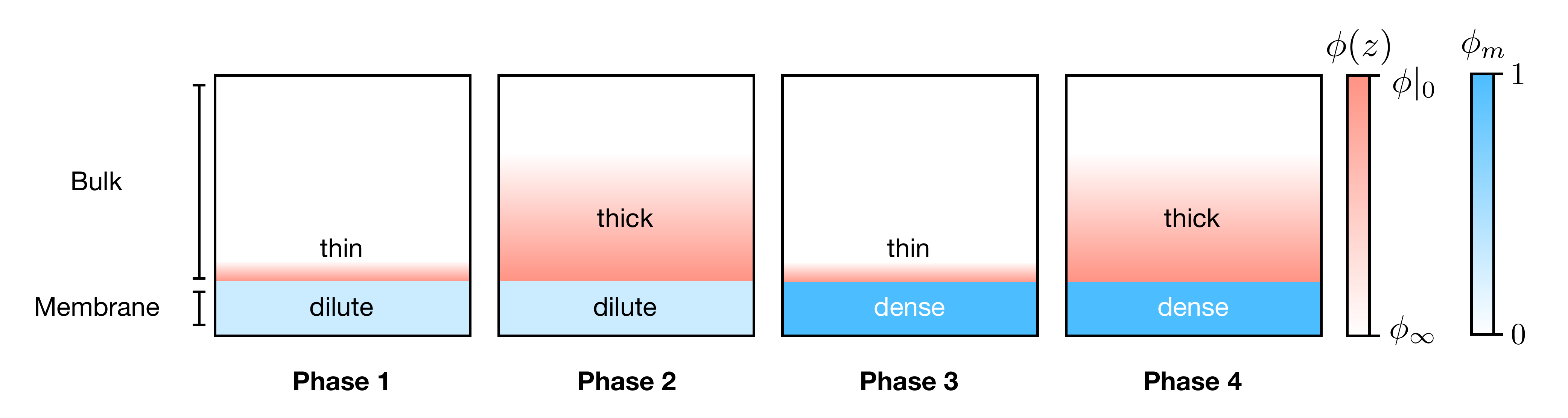}}
\caption{\textbf{Overview of surface states for systems where molecules can bind to a membrane surface.} We find four surface states which are each characterized by a pair of order parameters, i.e., the volume fraction of molecules bound to the surface, $\phi_m$, and  the excess surface concentration, $c_s$ (Eq.~\eqref{eq:surface_excess_conc}).
Depending on the interactions between bulk and surface, the membrane can be either dilute (light blue, low $\phi_m$) or dense (dark blue, high $\phi_m$), and the bulk layer can be either thin (low $c_s$) or thick (high $c_s$).
The shades of red show
the bulk profile $\phi(z)$
which ranges between the bulk volume fraction $\phi_{\infty}$ and the  volume fraction at the membrane surface $\phi\vert_0$ (see colorbar). 
}
 \label{fig:phases_sketch}
 \end{figure}

Due to the significance of both binding processes and bulk condensation in living cells we study  {how membrane binding affects wetting, prewetting and surface phase transitions.} To approach this question, we derive the corresponding thermodynamic theory in the presence of binding processes of molecules between bulk and membrane surfaces.
Our theory goes beyond the classical thermodynamics of phase transitions at and adjacent to surfaces by introducing new surface states.
Recently, related questions were addressed
using Monte-Carlo simulations of a ternary lattice model for mobile tethers that are confined in a membrane and that can bind bulk molecules~\cite{Rouches2021.02.17.431700}. 
This work focuses on the role of three phase coexistence in the membrane.

In our theory we focus on a binary mixture in the bulk that interacts with a surface layer bound to a flat and rigid membrane.
Our  thermodynamic theory shows that membrane binding can lead to a variety of thermodynamic surface states at undersaturated conditions with unexpectedly rich surface phase diagrams (see Fig.~\ref{fig:phases_sketch}).
Such states are described by a pair of order parameters, i.e., the fraction of bound molecules in a single surface layer and the excess surface concentration of the condensates adjacent to the surface.
Interestingly, we find cases where phase transitions at and adjacent to surfaces occur under conditions where the bulk cannot phase-separate at any concentration. 
More generally, a layer of bound molecules on the membrane effectively modifies the properties of the surface which can for example lead to a shift of the prewetting line to low concentrations. 
Finally, surface binding affects the wetting transition and the contact angle of bulk droplets that wet the surface. 
 
Our findings additionally suggest that the binding of molecules provides a versatile mechanism to control the position of wetted droplets  {at phase coexistence}. 
In recent years, a growing number of intra-cellular condensates were shown to adhere to membrane-bound organelles or the intra-cellular surfaces~\cite{Gall1999,
Brangwynne2009, Oliver_Cell_2019,Wenwenyu2020,
Knorr_wetting_nature_2021}.
Moreover, many of such condensates are suggested to act as scaffolds for biochemical processes~\cite{boeynaems2018protein, alberti2019considerations}.
Specific binding receptors at the  membrane surface can control the position of such wetted droplets and thereby spatially control  biological processes occurring inside droplets.

 {We also find that binding alters the prewetting behavior and the occurrence of surface phase transitions at concentrations below saturation.}
 {Since molecules can bind to the membrane surfaces, phase-separation can occur in the membrane (surface phase transition) altering the prewetting behavior by effectively modifying the  properties of the surface.
Moreover, while the classical prewetting transition line is very close to the saturation concentration without binding processes~(\cite{Schmidt_Moldover1983,Bonn1992}; see also Fig.~\ref{fig:coupling_shift}(a)),
the transition lines of prewetting and surface phase transitions can shift to lower values when molecules can bind to membrane surfaces. }
Interestingly, 
the actual physiological concentrations of  many membrane-binding proteins in living cells
(typically $(10-100) nM$~\cite{Oliver_Cell_2019,Goehring2011, banjade2014phase, su2016phase}) are far below their saturation concentrations (typically $(1-10)\mu M$~{\cite{Oliver_Cell_2019,Wenwenyu2020}}). 
Further research is required to scrutinize whether the low physiological concentrations of membrane-binding proteins serve the purpose to form  condensates on intracellular surfaces rather than droplets in the bulk.

Returning to Pauli's famous quote, surfaces in biological systems were probably not invented by the devil. 
Rather, membranes and surfaces have an essential functional role for living cells. 
This essential role is reflected in a plethora of biological processes ranging from cell division to intra- and extra-cellular transport. Binding to such surfaces together with surface phase transitions gives rise to
a new level of complexity that is exemplified in the rich variety of phases and variability of phase diagrams revealed by our work. This suggests that this additional complexity could play a key role in cell biological processes. 
We expect that this complexity is further extended 
when  {orientations of surface bound molecules can give rise to nematic phases~\cite{milchev2021cylindrical}}.
 {Complex patterns at surface are expected 
when binding processes are maintained away from thermodynamic equilibrium, which is the case in living cells.}
In biological systems,  ATP-driven cycles of kinase and phosphatase can alter binding equilibria. Future research will clarify how such active binding processes modify the properties of wetted and prewetted states out of equilibrium.

\acknowledgments{We thank S.\ Liese, J.\ Bauermann, L.\ Hubatsch, S. Bo, S. Laha, T.\ Harmon, I.\ LuValle-Burke, and D.\ Sun for insightful discussions and S. Liese for helpful feedback on the manuscript.
G.\ Bartolucci, A.\ Honigmann, and C.\ Weber acknowledge the SPP 2191
``Molecular Mechanisms of Functional Phase Separation'' of the German Science Foundation for financial support and for providing an excellent collaborative environment.
}
\bibliography{main}{}

\clearpage

\appendix

\section{Graphical construction for surface phase transitions with membrane binding}\label{APP:graphical_construction}

\begin{figure}
\centering
{\includegraphics[width=1.0\textwidth]{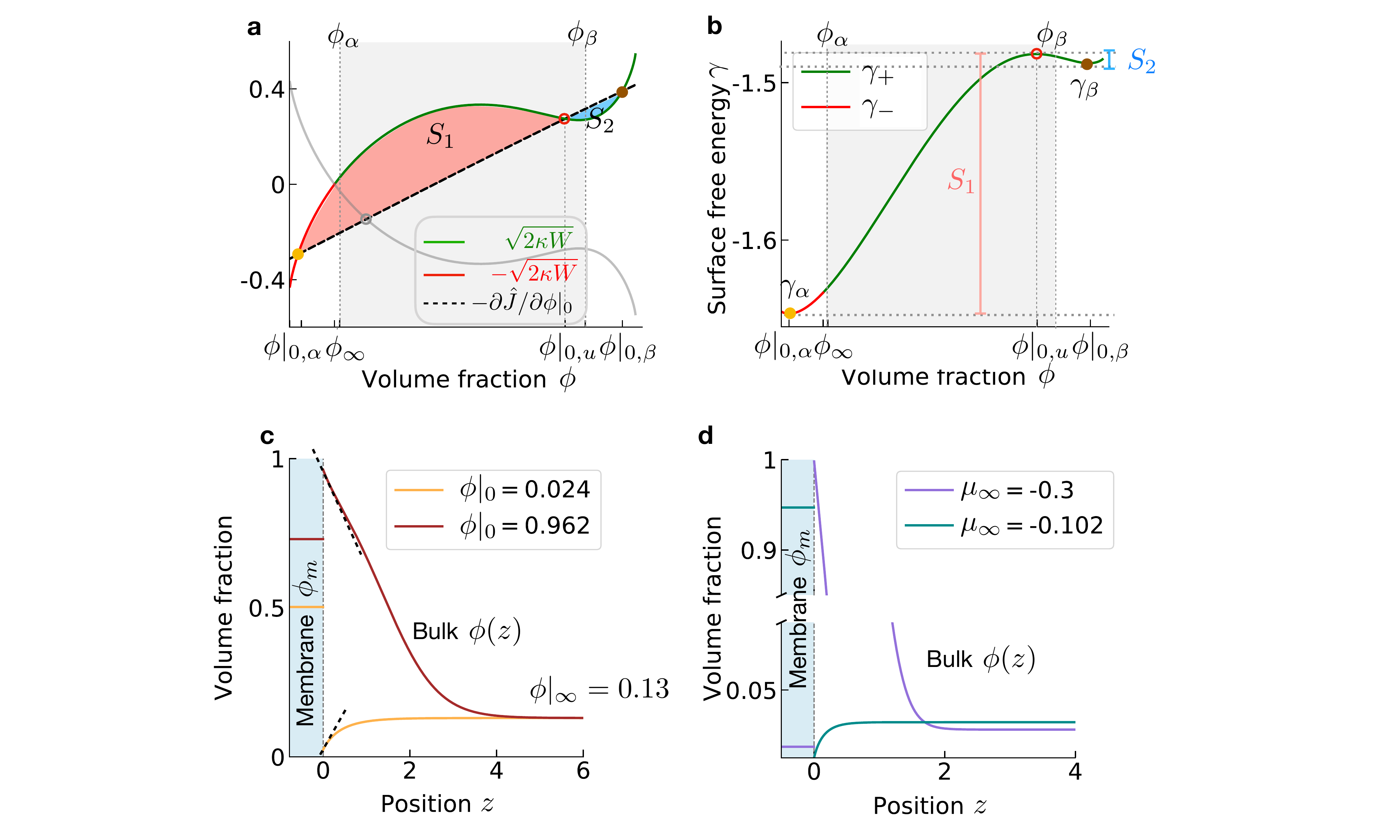}}
% graphi_construction_both_signs.pdf
\caption{\textbf{Graphical construction for wetting with both solution branches in Eq.~\eqref{eq:intersection_eqn2}.}
\textbf{(a)}
$-{\partial \hat{J}(\phi, \phi_m)}/{\partial \phi} = - \kappa \, d\phi/dz$ (dashed line) can intersect with both branches $\sqrt{2\kappa W}$  (solid orange line) and $-\sqrt{2\kappa W}$  (solid blue line).
 {Please note that we depict the non-physical branches with light grey curves in figure (a). 
The intersection points with with the non-physical branches  (grey empty circle(s)) do not lead to physical solutions because  
$(\phi_{\infty} - \phi\vert_0)\, d\phi/dz\vert_0 < 0$ and therefore the boundary condition $\phi(z)|_{z\to \infty}=\phi_\infty$ cannot be satisfied. 
The intersection points correspond to two local minima (solid yellow and brown points) and  {one} local maxima (empty red points).} 
\textbf{(b)} We create a free energy landscape $\gamma$ 
by varying the surface volume fraction while maintaining one of the equilibrium conditions Eq.~\eqref{eq:finale_EQ_condd}. 
For each branch in (a), we calculate the surface free energy $\gamma_+ = \int_{\phi_{\infty}}^{\phi} d\phi^\prime \Big[ \sqrt{2 \kappa W(\phi^\prime) }  + {\partial \hat{J}}/{\partial \phi^\prime} 
   \Big] $ $ + \hat{J}\left(\phi_{\infty}, \phi_m \right)$ (green line) and $\gamma_- = \int_{\phi_{\infty}}^{\phi} d\phi^\prime \Big[ - \sqrt{2 \kappa W(\phi^\prime) } +
   {\partial \hat{J}}/{\partial \phi^\prime} 
   \Big] + \hat{J}\left(\phi_{\infty}, \phi_m \right)$ (red line). The stability of the solutions (solid and empty symbols) is  determined by the local curvature of Gibbs surface free energy.
The gray shaded area represents the coexistence regime where the bulk can phase separation into a dilute phase with volume fraction $\phi_{\alpha}$ and a dense phase where the volume fraction is $\phi_{\beta}$.
 {\textbf{(c)} Concentration profiles of the two locally stable solutions: the first solution with $\phi\vert_0 = 0.024$ and positive slope (i.e. ${d\phi}/{dz} > 0$) nearby the surface (solid yellow line), while brown solid curve depicts the final locally stable solution with $\phi\vert_0 = 0.962$ and negative slope (i.e. ${d\phi}/{dz} < 0$). The dashed black line depicts the tangential lines of the concentration profiles at surface (i.e. $z = 0$).} \textbf{(d)}  {Two examples of concentration profile with $\chi_b = 3.7$, chemical potential $\mu_{\infty} = -0.3$ and $\mu_{\infty} = -0.102$, respectively in Fig.~\ref{fig:slaved_phase_diagram}d}: one with convex shape nearby the surface (purple solid line), while another with concave meniscus (green solid line).
}
\label{fig:graphic_method_both_signs}
\end{figure}

Eqs.~\eqref{eq:EQ_cond} are one differential equation of second order for the bulk volume fraction $\phi(z)$ coupled to three algebraic equations at the membrane boundary and at $z\to \infty$. In this section, we integrate once, leading to a differential equation of first order coupled to two algebraic equations. The resulting set of equations can then be solved by a graphical construction, a procedure which is similar to Ref.~\cite{Cahn_wetting_1977}.

After multiplying Eq.~\eqref{eq:EQ_conda} by $({\partial \phi}/{\partial z})$ on both sides, one obtains
\begin{align}\label{eq:EQ-cond1}
     \frac{\partial f_{b}}{ \partial z}  -\frac{1}{\nu_b} \mu_\infty \frac{\partial \phi}{\partial z} -  \frac{\partial}{\partial z} \left( \frac{\kappa}{2} \left(\partial_z \phi\right)^2 \right) = 0 \, .
\end{align}
Integrating this equation over the bulk,
\begin{align}
     \int_{z}^{\infty} dz' \left[ \frac{\partial f_{b}}{ \partial z'}  - \frac{1}{\nu_b} \mu_\infty \frac{\partial \phi}{\partial z'} -  \frac{\partial}{\partial z'} \left( \frac{\kappa}{2} \left(\partial_{z'} \phi\right)^2 \right) \right] = 0 \, ,
\end{align}
and utilising boundary conditions 
$ \partial_z \phi \vert_{\infty}= 0$ (Eq.~\eqref{eq:EQ_condb}),
and $\phi(\infty)=\phi_\infty$, we obtain
\begin{align}
\label{eq:integrated}
     0&=  f_{b}\left(\phi(z) \right) - f_{b} (\phi_{\infty})  - \mu_\infty \left( \frac{1}{\nu_b}\phi(z) - \frac{1}{\nu_b}\phi_{\infty}\right) - \frac{\kappa}{2} \left(\partial_z \phi \right)^2  \, .
 \end{align}
Using the definition of $W(\phi)$ (Eq.~\eqref{eq:def_W}),
 Eq.~\eqref{eq:integrated} can be written as
 \begin{align}
    |\partial_z \phi|^2 =  {\frac{2}{\kappa} W(\phi)} \, ,
\end{align}
leading to two solution branches (orange and blue lines in Fig.~\ref{fig:graphic_method_both_signs}):
\begin{align}\label{eq:intersection_eqn2}
    \partial_z \phi = \pm \sqrt{\frac{2}{\kappa} W(\phi)} \, .
\end{align}
Finally, substituting these two relations into Eq.~\eqref{eq:EQ_condc}, we obtain Eq.~\eqref{eq:finale_EQ_condc}. Please note that since we have already used Eq.~\eqref{eq:EQ_condb} in the derivation above, we are left with two algebraic equations and one ODE of first order (see Eqs.~\eqref{eq:finale_EQ_cond}).
These equations can be solved via a graphical construction 
to obtain the surface volume fraction $\phi|_0$ and membrane area fraction $\phi_m$ at thermodynamic equilibrium; see Fig.~\ref{fig:graphic_method_both_signs} for an illustration in the presence of  the two solution branches (Eq.~\eqref{eq:intersection_eqn2}).

Numerical methods to obtain solutions to Eqs.~\eqref{eq:finale_EQ_cond} may suffer from the fact that 
the bulk chemical potential $\mu_b = {\partial f_b}/{\partial \phi}$ is singular for $\phi = 1$.
To avoid numerical problems related to this singularity, we can add a term of the form $- \epsilon (1-\phi) \ln(1-\phi)$ to the coupling free energy $J$ such that $-{\partial J}/{\partial \phi\vert_0}$ can always intersect with $\sqrt{2 \, \kappa \, W}$ in the limit $\phi \to 1$. 
We used $\epsilon={10^{-3}}$, which is a very small number to ensure that the thermodynamic states are hardly affected. 
%For simplicity, we assume $\phi = 0.999$ as always our solution of eqn.~\eqref{eq:graphi_construct_eq} instead of adding this nonlinear term to the coupling free energy $J$.

\section{Condition of critical prewetting}\label{APP:analytical_deri_prewetting_critical_point}

In this section, we derive the condition for prewetting critical points with membrane binding. 
At the prewetting critical point, the Gibbs surface potential $\gamma_s(\phi\vert_0)$ given in Eq.~\eqref{eq:Gibbs_energy_EQ} has
a zero curvature with respect to $\phi\vert_0$, 
\begin{align} \label{eq:second_order_deriv}
    \frac{\kappa W'(\phi\vert_0) }{\sqrt{2 \kappa W(\phi|_0) }}\pm \frac{\partial^2 J(\phi, \phi_m)}{\partial \phi^2}\Big|_0 \pm \frac{\partial^2 J(\phi, \phi_m)}{\partial \phi \, \partial \phi_m} \frac{\partial \phi_m}{\partial \phi}\Big|_0  &= 0 \, ,
\end{align}
where ${\partial \phi_m}/{\partial \phi}\vert_0$ can be obtained from the derivative of  Eq.~\eqref{eq:finale_EQ_condd}: 
\begin{align}
 \frac{\partial^2 f_m}{\partial \phi_m^2} \frac{\partial \phi_m}{\partial \phi}\Big|_{0}  + \frac{\partial^2 J(\phi, \phi_m)}{\partial \phi_m^2} \frac{\partial \phi_m}{\partial \phi}  \Big|_{0}  +   \frac{\partial^2 J(\phi, \phi_m)}{\partial \phi_m \partial \phi}   \Big|_{0}& = 0 \, .
\end{align}
Substituting $J(\phi, \phi_m)$ given in Eq.~\eqref{eq:J_definition} and $f_m$ given in Eq.~\eqref{eq:free_energy}, we obtain 
\begin{align}\label{eq:dphim_dphi}
    \frac{\partial \phi_m}{\partial \phi}\Big|_{0} = \chi_{bm} \frac{\phi_m (1-\phi_m)  }{(1 - 2\chi_m \phi_m (1 - \phi_m))} \, . 
\end{align}
Combining Eq.~\eqref{eq:second_order_deriv} with Eq.~\eqref{eq:dphim_dphi}, we find the condition for prewetting critical point,
% \begin{align}
% \frac{\kappa \frac{k_B T}{\nu}\Big [  \frac{1}{n_b} \ln \phi  - \ln \left(1-\phi\right) + \chi_b \, (1- 2 \,\phi) \Big] -\kappa \frac{\mu_{\infty}}{\nu_b} }{\sqrt{2 \kappa W(\phi)}}\Bigg\vert_0 \pm \\ \pm \frac{k_B T}{\tilde{\nu}}\chi_{bb} \pm \frac{k_B T}{\tilde{\nu}} \chi_{bm}^2  \frac{\phi_m (1-\phi_m)  }{(1 - 2\chi_m \phi_m (1 - \phi_m))}  &= 0 \, ,
% \end{align}
\begin{align}
\sqrt{\frac{ \kappa }{2 W(\phi)}}  \frac{\tilde{\nu}}{\nu_b} \Bigg[  \ln \phi  - n_b\ln \left(1-\phi\right) + n_b \chi_b \, (1- 2 \,\phi)  - \frac{\mu_{\infty}}{\, k_B T} \Bigg]_0  \nonumber
 \\ \pm  \left[ \chi_{bb} + \chi_{bm}^2  \frac{\phi_m (1-\phi_m)  }{(1 - 2\chi_m \phi_m (1 - \phi_m))}  \right] &= 0 \, ,
\end{align}
where $W(\phi)$ is defined in Eq.~\eqref{eq:def_W}.

 {
\section{Comparison with Nakanishi and Fisher's results}\label{APP:comparision_Nakanishi_Fisher}}

 {In this section, we compare our results with Nakanishi and Fisher's results~\cite{Nakanishi_PRL_1982}. 
We distinguish two scenarios in the following discussion: with and without phase separation in the membrane-bound layer.} 

\begin{figure}
\centering
{\includegraphics[width=1.0\textwidth]{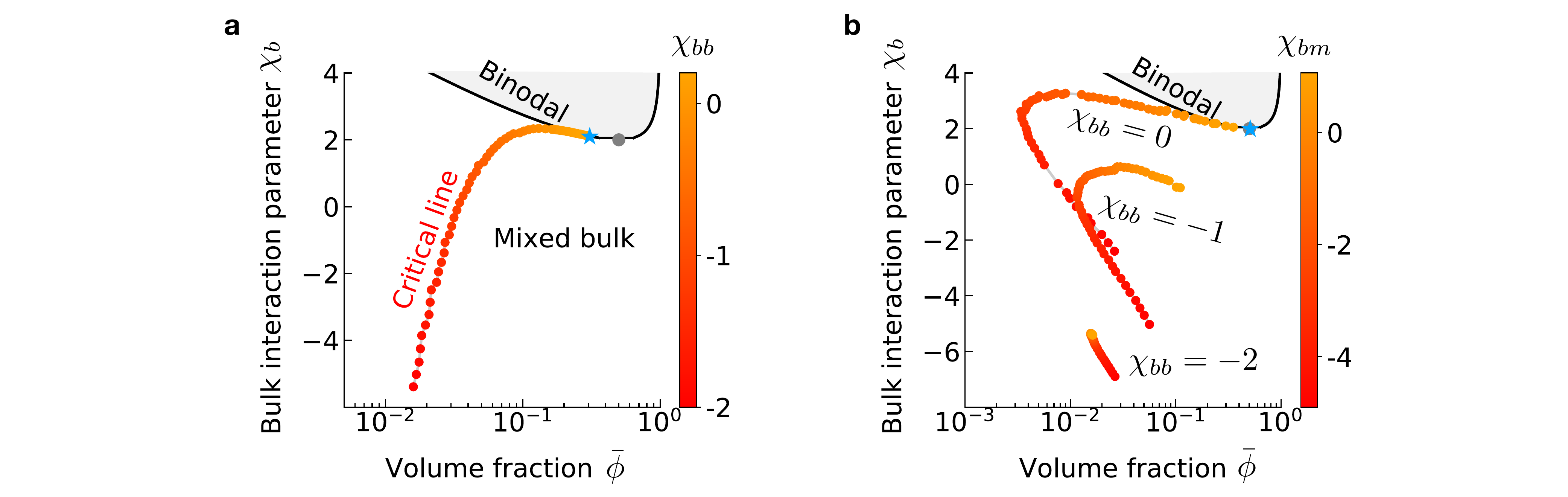}}
\caption{
 {\textbf{Prewetting critical points.}  \textbf{(a)} Without binding to the membrane surface, prewetting critical points decrease in both bulk interaction parameter $\chi_b$ and bulk volume fraction $\bar \phi$ as the surface enhancement parameter $\chi_{bb}$ decreases. Critical wetting occurs when the prewetting critical point merges with the binodal (blue star).
\textbf{(b)} With surface binding,
prewetting critical points can have a minimum bulk volume fraction $\bar \phi$ as the coupling between membrane and bulk gets more attractive ($\chi_{bm}$ decreases).
For non-zero and more pronounced surface enhancement  (decreasing $\chi_{bb}$), the critical points can no more merge the binodal suppressing the critical wetting. 
Note that the line of prewetting critical points only terminate  because we limited our study to a finite range in $\chi_{bm} \in [-4.8,1]$. 
Parameter values for (a) and (b) are: $\chi_m = -4$, $\omega_b = -0.3$.}}
\label{fig:comparison}
\end{figure}

 {If the membrane-bound layer cannot phase-separate (i.e., $\chi_m < 2$ for $n_m=1$), the terms in the coupling free energy, $\chi_{bm} 
\phi_m \phi_s$, and $\chi_{bb} \phi^2$ can both act as an effective surface enhancement for $\chi_{bm} < 0$ and $\chi_{bb} < 0$.
Here, we scrutinize whether the effects of the coupling parameter $\chi_{bm}$ are indeed qualitatively similar to the ones of the classical enhancement parameter $\chi_{bb}$ as analyzed previous models~\cite{Cahn_wetting_1977, Nakanishi_PRL_1982}.
In Fig.~\ref{fig:comparison}, we show a $\chi_b-\bar \phi$ phase diagrams and depict  the critical prewetting points, (a) for the case without binding when varying $\chi_{bb}$, and 
(b) for the case with membrane binding as a function of the coupling parameter $\chi_{bm}$ for $\chi_{bb}=0$. 
We find that both cases share a similarity, namely when decreasing $\chi_{bb}$ for the case without binding or decreasing $\chi_{bm}$ for the case with binding, the prewetting points shift to lower bulk interaction parameters $\chi_b$.
However, both cases also show qualitative differences. 
First, for the case without binding, the critical points shift to lower bulk volume fraction $\bar \phi$, while there is a minimum in $\bar \phi$ for the case with binding. 
Second, without binding, there is a $\chi_{bb}$ value where the prewetting critical point coincides with the binodal line implying a continuous wetting transition as reported in Ref.~\cite{Nakanishi_PRL_1982}. 
This is different to the case with surface binding and $\chi_{bb}=0$, where the prewetting critical point merges with the bulk critical point, thereby fusing bulk and surface criticality. 
For non-vanishing surface enhancement, e.g., $\chi_{bb}=-1$, critical wetting is even suppressed for the case with binding since the critical prewetting line remains disconnected from the binodal for any coupling parameters $\chi_{bm}$. 
% for attractive coupling parameters $\chi_{bm}<0$.
In summary, in the absence of phase separation in the membrane, the case with and without binding leads to a qualitative
different thermodynamic behavior at surface. 
}

%%%% about continuous wetting trasniton
%If $\chi_{bb}=0$, there is cont. wetting transitions in our model with membrane binding only at the bulk critical point (prewetting critical point meet the binodal and not at the bulk critical point). However, $\chi_{bm} > 0$ can not lead to the critical wetting, which is achieved by $\chi_{bb}>0$.  The reason is that the condition of  critical wetting is  ${\partial^2 J}/{\partial \phi\vert_0^2} = - {\partial \sqrt{2\kappa W}}/{\partial \phi\vert_0} > 0$ (see Fig.~\ref{fig:critical_wetting}a). With surface binding but set $\chi_{bb} = 0$, due to ${\partial^2 J}/{\partial \phi\vert_0^2} = \chi_{bm} {\partial \phi_m}/{\partial \phi_s} < 0$ in the globally stable branch (we show the mathematical proof at the end of this section), it is impossible to achieve critical wetting. 

 {
If the membrane-bound layer can phase-separate (i.e., $\chi_m > 2$ for $n_m=1$), the case with and without binding leads even more qualitative differences. 
For example, 
we obtain multiple surface states (see Fig.~\ref{fig:multiple_preweeting1}a and Fig.~\ref{fig:multiple_preweeting2}a), which can not be obtained from Nakanishi and Fisher's results.
Furthermore, if we consider a relationship between the interaction parameters $\chi_{bm}$ and $\chi_b$ (e.g., Eq.~\ref{eq:chibm_chib}, we find prewetting transition lines which are not connected with coexistence line (see Fig.~\ref{fig:slaved_phase_diagram}(a-b)).
}

%\begin{figure}
%\centering
%{\includegraphics[width=1.0\textwidth]{./Figure11_rebuttal.pdf}}
%\caption{\textbf{Critical wetting illustration.} \textbf{(a)} Graphic construction condition at critical wetting transition. The slope of the dash line $-{\partial J}/{\partial \phi\vert_0}$ is negative. In the previous model without surface binding \cite{Cahn_wetting_1977,Nakanishi_PRL_1982}, we can get critical wetting with ${\partial J}/{\partial \phi\vert_0} = \chi_{bb} > 0$. In our model, if we set $\chi_{bb} \le 0$, it's impossible to achieve continuous critical wetting, since ${\partial^2 J}/{\partial \phi\vert_0^2} = \chi_{bm} {\partial \phi_m}/{\partial \phi\vert_0} <0$, see Eq.~\eqref{eq:coupling_second_order}.  \textbf{(b)} For each branch in (a), we calculate the surface free energy $\gamma_+ = \int_{\phi_{\infty}=\phi_{\beta}}^{\phi} d\phi^\prime \Big[ \sqrt{2 \kappa W(\phi^\prime) }  + {\partial \hat{J}}/{\partial \phi^\prime}   \Big] $ $ + \hat{J}\left(\phi_{\infty}, \phi_m \right)$ (green line) and $\gamma_- = \int_{\phi_{\infty}=\phi_{\beta}}^{\phi} d\phi^\prime \Big[ - \sqrt{2 \kappa W(\phi^\prime) } +   {\partial \hat{J}}/{\partial \phi^\prime}    \Big] + \hat{J}\left(\phi_{\infty}, \phi_m \right)$ (red line). Two Gibbs surface free energy branches merges at dense phase $\phi_{\beta}$ and second order derivative of green curve at this point equals to zero.}
%\label{fig:critical_wetting}
%\end{figure}

\end{document}